%% file: ms.tex
\pdfoutput=1

\documentclass[conference]{IEEEtran}
\IEEEoverridecommandlockouts
\usepackage[utf8]{inputenc}
\usepackage[english]{babel}
\usepackage{array}
\usepackage{multirow} 
\usepackage{graphicx}
\usepackage{url}
\usepackage[%
comma,         
numbers,       
]{natbib} 
\usepackage{environ}
\usepackage{siunitx}
\usepackage{placeins}
\usepackage{tabularx, ragged2e}
\usepackage[most]{tcolorbox}

\usepackage{enumitem}
\usepackage{booktabs}
\usepackage{float}
\usepackage{caption} 
\usepackage{newfloat}
\usepackage{booktabs}
\usepackage{glossaries}
\usepackage{cleveref}
\usepackage{multicol, blindtext}
\usepackage{listings}
\usepackage{makecell}
\usepackage{microtype}

\input{definitions}

\begin{document}

\maketitle

\sloppy

\input{abstract}
\input{introduction}

\input{background.tex}

\input{adversaries.tex}

\input{design_goals.tex}

\input{oscore.tex}

\input{edhoc.tex}

\input{evaluation.tex}

\input{related_work}

\input{conclusion.tex}

\bibliographystyle{IEEEtranS.bst}
\bibliography{bibliography}

\end{document}

%% file: definitions.tex
\title{The Cost of OSCORE and EDHOC for \\ Constrained Devices\\
{\normalsize \textit{Extended Paper \textsuperscript{\S}}}
\thanks{\textsuperscript{\S} A short version of this paper will appear on CODASPY 2021.}
}

\author{
 	\IEEEauthorblockN{
	Stefan Hristozov\IEEEauthorrefmark{1},
	Manuel Huber\IEEEauthorrefmark{2},
	Lei Xu\IEEEauthorrefmark{1},
	Jaro Fietz\IEEEauthorrefmark{1},
	Marco Liess\IEEEauthorrefmark{1},
 	Georg Sigl\IEEEauthorrefmark{3}}
	
 	\IEEEauthorblockA{\IEEEauthorrefmark{1}Fraunhofer AISEC, Garching Germany, 
 	\{stefan.hristozov, lei.xu, jaro.fietz, marco.liess\}@aisec.fraunhofer.de }
	
 	\IEEEauthorblockA{\IEEEauthorrefmark{2}Microsoft, Vancouver Canada,
	 manuel.huber@microsoft.com}
	 
	\IEEEauthorblockA{\IEEEauthorrefmark{3}TU Munich, Munich, Germany,
	sigl@tum.de}
 }

\newacronym{tls}{TLS}{Transport Layer Security}
\newacronym{dtls}{DTLS}{Datagram Transport Layer Security}
\newacronym{ietf}{IETF}{Internet Engineering Task Force}
\newacronym{coap}{CoAP}{Constrained Application Protocol}
\newacronym{oscore}{OSCORE}{Object Security for Constrained RESTful Environments}
\newacronym{edhoc}{EDHOC}{Ephemeral Diffie-Hellman Over COSE}
\newacronym{rest}{REST}{Representational State Transfer}
\newacronym{uri}{URI}{Uniform Resource Identifier}
\newacronym{tee}{TEE}{Trusted Execution Environment}
\newacronym{wot}{WoT}{Web of Things}
\newacronym{psk}{PSK}{Pre-Shared Key}
\newacronym{rpk}{RPK}{Raw Public Key}
\newacronym{cert}{CERT}{Certificate}
\newacronym{aad}{AAD}{Additional Authenticated Data}
\newacronym{aead}{AEAD}{Authenticated Encryption with Associated Data}
\newacronym{pfs}{PFS}{Perfect Forward Secrecy}
\newacronym{cbor}{CBOR}{Concise Binary Object Representation}
\newacronym{cose}{COSE}{\gls{cbor} Object Signing and Encryption}
\newacronym{ecdh}{ECDH}{Elliptic Curve Diffie-Hellman}
\newacronym{dh}{DH}{Diffie-Hellman}
\newacronym{prk}{PRK}{Pseudorandom Key}
\newacronym{ikm}{IKM}{Input Keying Material}
\newacronym{okm}{OKM}{Output Keying Material}
\newacronym{mac}{MAC}{Message Authentication Code}
\newacronym{ble}{BLE}{Bluetooth Low Energy}
\newacronym{spm}{SPM}{Secure Partition Manager}
\newacronym{spu}{SPU}{System Protection Unit}
\newacronym{hkdf}{HKDF}{HMAC-based Key Derivation Function}
\newacronym{ca}{CA}{Certification Authority}
\newacronym{sw}{SW}{Secure World}
\newacronym{nsw}{NSW}{Non Secure World}
\newacronym{cost}{COST}{Commercial Off-The-Shelf}
\newacronym{os}{OS}{Operating System}
\newacronym{rtos}{RTOS}{Real-Time \gls{os}}
\newcommand{\uoscore}{\textmu OSCORE}
\newcommand{\uedhoc}{\textmu EDHOC}
\newcommand{\uoscoretee}{\textmu OSCORE-TEE}
\newcommand{\uedhoctee}{\textmu EDHOC-TEE}

%% file: abstract.tex
\begin{abstract}
    Many modern IoT applications rely on the \gls{coap} because of its efficiency and seamless integrability in the existing Internet infrastructure. One of the strategies that \gls{coap} leverages to achieve these characteristics is the usage of \textit{proxies}.
    Unfortunately, in order for a proxy to operate, it needs to terminate the (D)TLS channels between clients and servers. Therefore, end-to-end confidentiality, integrity and authenticity of the exchanged data cannot be achieved. 
    In order to overcome this problem, an alternative to (D)TLS was recently proposed by the \gls{ietf}. This alternative consists of two novel protocols: 1) \gls{oscore} providing authenticated encryption for the payload data and 2) \gls{edhoc} providing the symmetric session keys required for \gls{oscore}.
    In this paper, we present the design of four firmware libraries for these protocols especially targeted for constrained microcontrollers and their detailed evaluation. 
    More precisely, we present the design of \uoscore{} and \uedhoc{} libraries for regular microcontrollers and \uoscoretee{} and  \uedhoctee{} libraries for microcontrollers with a \gls{tee}, such as microcontrollers featuring ARM TrustZone-M. 
    Our firmware design for the later class of devices concerns the fact that attackers may exploit common software vulnerabilities, e.g., buffer overflows in the protocol logic, OS or application to compromise the protocol security. 
    \uoscoretee{} and  \uedhoctee{} achieve separation of the cryptographic operations and keys from the remainder of the firmware, which could be vulnerable. 
    We present an evaluation of our implementations in terms of RAM/FLASH requirements, execution speed and energy on a broad range of microcontrollers. 
    %
\end{abstract}

%% file: introduction.tex
\section{Introduction}
\gls{coap} is a widely used IoT application layer protocol. It was designed with two main goals: 1) to be efficient for large deployments of constrained devices communicating over constrained networks and 2) to integrate easily with existing \gls{rest} \cite{Fielding2000} protocols, such as HTTP \cite{rfc7230, rfc7252, rfc8075}. 
One of the strategies that \gls{coap} uses to achieve these goals is to leverage proxy software executed on middle boxes, such as the border routers connecting the constrained IoT networks and the non-constrained networks, e.g., the Internet \cite{Sulaeman2016, Ludovici2015}.
For example, a proxy may serve incoming requests from a cache containing previously received data that is still valid in order to save the constrained communicational and computational resources of the device.
In other use cases, the proxy may act as a \textit{cross-protocol proxy} and translate between \gls{coap} and HTTP.
Such cross-protocol proxies achieve integration of the constrained devices on the Internet that is independent of the application logic \cite{rfc8075, Castellani2012, Lerche2012}.

The \gls{coap} specification \cite{rfc7252} states that DTLS or TLS \cite{rfc8323, rfc7925} has to be used to secure \gls{coap}.
Although (D)TLS provides strong security guaranties, (D)TLS channels must be terminated at the proxies. This is because the proxies need to access certain fields of the \gls{coap} or the HTTP message, which are encrypted at the transport layer. Decrypting the secured messages at the proxies makes the necessary fields available, but unfortunately also exposes the sensitive message payload \cite{Selander2017, rfc8613}. 

In order to circumvent this security issue, an alternative to (D)TLS consisting of the protocols \gls{oscore} \cite{rfc8613} and \gls{edhoc} \cite{edhoc} was recently proposed by \gls{ietf}. In contrast to (D)TLS, \gls{oscore} and \gls{edhoc} operate on the application layer, thus making proxy operations possible without decrypting the message payload.
In addition, in order for both protocols to be well suited for constrained devices and networks, they leverage the low overhead encoding formats \gls{cbor} \cite{rfc7049} and \gls{cose} \cite{rfc8152}.

Previous work in the area of \gls{oscore} and \gls{edhoc} implementations only considers early protocol drafts and therefore does not include all protocol modes \cite{Perez2019}. The previous work also does not provide detailed evaluation in terms of memory, execution speed and energy on constrained microcontrollers \cite{Gndoan2020}. We, however, describe the design of our libraries and provide the first detailed evaluation of the latest states of the specifications \cite{rfc8613,edhoc}. Moreover, we consider all modes of operation.   
Furthermore, previous work does not consider the fact that IoT devices are often prone to software vulnerabilities, e.g., buffer overflows, which may compromise even the best-designed protocols. 
This is due to the fact the IoT devices often run firmware written in system languages such as C and C++. Moreover, such IoT firmware is often executed without an \gls{os} or on top of a \gls{rtos} which provides very limited or even no security features. An attacker exploiting such a software vulnerability could, for example, leak the cryptographic keys used in the protocols and use them to impersonate the device. We address this issue by identifying the critical parts of the \gls{oscore} and \gls{edhoc} protocols and placing them in a \gls{tee}.
Such \glspl{tee} have become recently available on main stream microcontrollers, for instance microcontrollers based on the Cortex-M23/33 cores featuring ARM TrustZone-M \gls{tee} \cite{M2351, nrf91, STM32L55, LPC5500}.
\subsection*{Contributions}
We present the design of the firmware libraries \uoscore{} and \uedhoc{} for regular microcontrollers. Our firmware designs consider all modes of operation of the final version of the \gls{oscore} specification \cite{rfc8613} and the latest \gls{edhoc} draft version \cite{edhoc}. 
%
%
The main feature of our designs is that they are completely independent of the \gls{coap} library, embedded OS, the communication protocol stack and the crypto library. \uoscore{} and \uedhoc{} are available as open source software \cite{opensourcelink}.

We present the design of the firmware libraries \uoscoretee{} and \uedhoctee{} for microcontrollers featuring a \gls{tee}. These libraries separate the cryptographic keys and routines from the rest of the firmware, which could be vulnerable. We designed \uoscoretee{} and \uedhoctee{} especially for microcontrollers, but they can also be ported to more powerful computing environments, e.g., ARM Cortex A class of devices.

We provide the first detailed study about the applicability of the \gls{oscore} and the \gls{edhoc} protocols on constrained IoT devices.
More precisely, we provide a detailed evaluation of our libraries regarding RAM/FLASH requirements and execution speed on four broadly used low-end CPUs: Cortex M0, M4, M33 and Xtensa.
We evaluate the overhead caused by using a \gls{tee} on the nRF9160 \cite{nrf91} radio SoC from Nordic Semiconductor featuring a Cortex-M33 CPU with TrustZone-M. 
In addition, we present an evaluation of the energy requirements of the protocols in an IPv6 over \gls{ble} network \cite{rfc7668}.

%% file: background.tex
\section{Background}
This section provides background for the rest of the paper. 
The \gls{coap} introduction provides details which are required for understanding the \gls{oscore} internals. The TrustZone-M introduction is required for understanding the separation of code and keys.   

\subsection{\gls{coap}}
\gls{coap} is a RESTful application layer protocol especially designed for the IoT domain \cite{rfc7252}. It considers two types of devices -- clients and servers, which communicate using requests and responses. The servers host \textit{resources} such as sensors and actuators. The clients may access those resources using PUT, GET, POST and DELETE methods. Each resource is reachable through a \gls{uri}. 
%
Each \gls{coap} packet begins with a fixed 4-byte header carrying the method type (PUT, GET, POST, DELETE) or a response code, among other information.
The header is followed by an optional token used to
correlate requests and responses.
The token is followed by optional options that contain additional parameters for the requests/responses.
These are followed by an optional payload, prefixed with the payload. 

\subsection{TrustZone-M}
A \gls{tee} is a state-of-the-art feature that has been available on application class processors for several years. Microcontroller class processors featuring a \gls{tee} called TrustZone-M were recently introduced by ARM with the ARMv8-M architecture. ARMv8-M chips such as nRF9160 and nRF5340 became recently available on the mass market. 

TrustZone-M separates a microcontroller in two domains called \textit{secure world} and \textit{non-secure world}. RAM/FLASH memory regions and memory-mapped peripherals of the microcontroller are associated with one of those domains. Secure world code, data and peripherals are not accessible from non-secure world code or peripherals. Secure world code and peripherals can access secure world and non-secure world code and peripherals.

Functions in the secure world can call functions in the non-secure world without any restrictions. Secure functions are called by non-secure functions using a special type of secure memory region called \textit{non-secure callable}. Only functions in the secure world defined with the non-secure callable attribute will contain a special \textit{secure gateway} (\texttt{SG}) instruction, which allows them to be called from the non-secure world. Such functions are referred to as veneer functions \cite{secureCode}. 
This is the only way in which secure software can be accessed from the non-secure software.  

%
%
By default, after power-up or reset the execution starts in the secure world. At that point, all memory regions are secure. The mapping of memories and peripherals into the secure world and the non-secure world is done during the boot process before the non-secure application is called.

%% file: adversaries.tex
\section{Adversarial Model}
This paper considers scalable remote software attacks targeting IoT devices. Such attacks may leverage common software vulnerabilities such as buffer overflows to leak cryptographic keys or manipulate the cryptographic operations involved in the \gls{oscore} and \gls{edhoc} protocols.    

However, we assume that the attacker cannot break state-of-the-art cryptographic algorithms or disturb their execution when they are executed inside the \gls{tee}. We also assume that the attacker cannot gain any knowledge of keys stored inside of the \gls{tee}.
This paper also does not consider physical attacks targeting the IoT devices such as physical side channels \cite{Lo2017}, fault injections e.g., glitching \cite{Obermaier2017} or physical memory dumping \cite{Obermaier2017a} since such attacks require physical access to the devices which makes them less scalable.

%% file: design_goals.tex
\section{Requirements and Design Goals}\label{sec:designgoals}
In this section, we state several high level requirements, which our firmware designs and their implementations have to fulfil. Then we derive a set of precise design goals from these requirements.

    \textbf{R-I: Lightweightness:} 
    \gls{oscore} and \gls{edhoc} are intended to be executed on constrained microcontrollers, e.g., microcontrollers based on the ARM Cortex M CPUs. Our firmware designs have to be suitable for this class of devices.  
    Therefore, the implementations should be fast, interruptible and have low FLASH/RAM requirements. Additionally, heap memory should not be used. Heap usage on a microcontrollers often leads to heap fragmentation, which in turn may lead to situation at runtime where no additional memory can be allocated \cite{heap}. 

    \textbf{R-II: Adoptability:} 
    Our firmware designs have to ensure the portability of the implementations to a variety of application environments, e.g., different \glspl{rtos}, radios and hardware platforms.
    Therefore, our firmware designs should not rely on any specific: 1) embedded OS, 2) CoAP library, 3) underlying protocol stack, 4) crypto library or 5) specific hardware features, e.g., crypto accelerators or radio transceivers. 

    \textbf{R-III: Firmware and Key Isolation:} 
    For our firmware designs for microcontrollers featuring a \gls{tee}, we assume that an attacker exists that can exploit software vulnerabilities such as buffer overflows, and in this way can compromise the security of the protocols. In order to mitigate such attacks, our designs for microcontrollers featuring a \gls{tee} should reduce the possibility of such software vulnerabilities being exploited, as much as possible. 
    Therefore, to achieve this requirement, the sensitive key material and crypto operations should be separated from the possibly vulnerable rest of the firmware. More precisely, only the bare minimum code should be placed in the \gls{tee}. In addition, the interface between the \gls{tee} code and the non-\gls{tee} code should be as narrow as possible.  


%% file: oscore.tex
\section{OSCORE}
In this section, we briefly describe the \gls{oscore} protocol functionality. Then we introduce the design of \uoscore{} and \uoscoretee.

\subsection{\gls{oscore} Functionality}\label{sec:oscorefunctionality}
\gls{oscore} secures \gls{coap} by providing end-to-end authenticated encryption, replay protection and binding of requests and responses. At the same time, \gls{oscore} allows proxy operations. 
More precisely, \gls{oscore} protects: 1) the method/response code, 2) the \gls{uri} of the requested resource and 3) the payload. 
An \gls{oscore} packet has the same structure as a \gls{coap} packet. The main difference between the two is that the payload of the \gls{oscore} packet is encrypted and integrity-protected by using an \gls{aead} algorithm \cite{Rogaway2002}. 
The \gls{aead} algorithm uses a shared symmetric keys on the client and the server, which may either be pre-established or established with a key establishment protocol such as \gls{edhoc} \cite{edhoc}.
\gls{oscore} packets are identified by the \gls{oscore} option field, which can also contain additional parameters. 

\subsubsection{\gls{oscore} Security Contexts}
Each \gls{oscore} endpoint maintains all the information it requires in three security contexts: a common context, a sender context and a recipient context. 
The common context is, as the name states, common to both client and server. It consists of identifiers for the \textit{\gls{aead} Algorithm} \cite{Rogaway2002} and the \textit{HKDF Algorithm} \cite{rfc5869}, \textit{Master Secret}, \textit{Master Salt}, \textit{ID Context} and \textit{Common IV}.
Algorithms that must be implemented are AES-CCM-16-64-128 \cite{rfc3610} and HMAC-SHA256 \cite{rfc5869}. 
\textit{Master Secret} is the shared secret. 
In addition to the common context, both \gls{oscore} client and server possess a sender and a recipient context. 
The parameters \textit{Sender ID} and \textit{Sender Key} of the sender contexts are identical to the parameters \textit{Recipient ID} and \textit{Recipient Key} of the recipient contexts. 
The \textit{Sender Key} and the \textit{Recipient Key} are symmetric keys derived from the \textit{Master Secret}. 
\textit{Sender ID} and \textit{Recipient ID} are identifiers used to identify the sender/recipient contexts.

\subsubsection{Conversion between \gls{coap} and \gls{oscore} and Vice Versa}\label{sec:oscore-coap conversion}
\begin{figure}[t] 
    \centering
    \includegraphics[width=\linewidth]{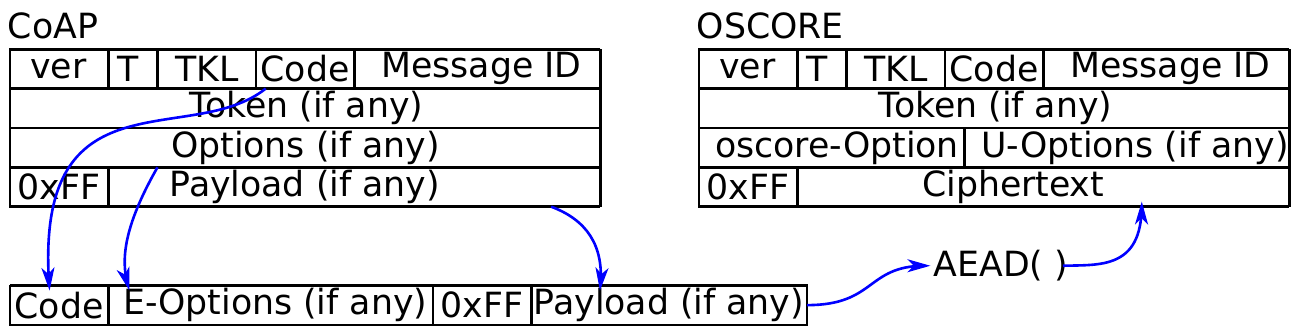}
    \caption{Converting \gls{oscore} message to a \gls{coap} message}
    \label{fig:coap-to-oscore}
\end{figure}

A \gls{coap} message is converted into an \gls{oscore} message before it is sent, and vice versa when it is received. 
\Cref{fig:coap-to-oscore} shows the conversion of a \gls{coap} message to \gls{oscore}.
The Code field, some options that need to be protected (called E options) and the optional payload form a plaintext, which is to be encrypted and integrity protected. The resulting cipher text is the payload of the \gls{oscore} packet. The Code of the \gls{oscore} packet is fixed -- 0.02 (POST) for requests and 2.04 (Changed) for responses. In addition, the \gls{oscore} packet contains an \gls{oscore} option field, which is used to distinguish an \gls{oscore} packet from a \gls{coap} packet.
Moreover, in some cases the \gls{oscore} option transports parameters used for the nonce generation and identification of the contexts at the receiving party.

\subsection{\uoscore{} Design}
\uoscore{} has a simple API, consisting of only three functions: 
\texttt{oscore\_init()}, \texttt{oscore2coap()} and \texttt{coap2oscore()}.
The \texttt{oscore\_init()} function initializes the \gls{oscore} contexts. The functions \texttt{oscore2coap()} and \texttt{coap2oscore()} convert \gls{oscore} to \gls{coap} packets and vice versa. 

\begin{figure}[t] 
    \centering
    \includegraphics[width=1\linewidth]{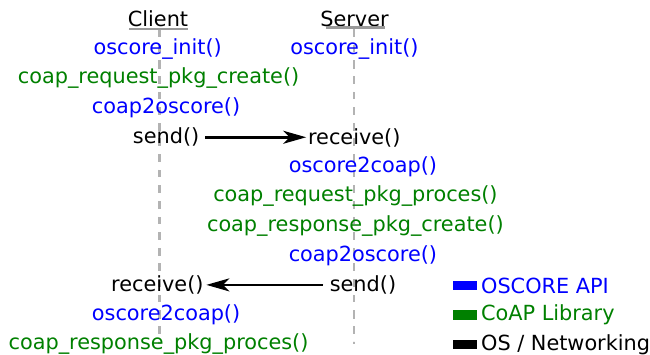}
    \caption{API usage model}
    \label{fig:usagemodel}
\end{figure}

\uoscore{} envisions a usage model in which the user's \gls{coap} application runs as usual using some \gls{coap} library and embedded OS preferred by the user. The conversion to/from \gls{oscore} happens just before a \gls{coap} packet needs to be sent and just after an \gls{oscore} packet is received, see \Cref{fig:usagemodel}.
The functions \texttt{coap\_request\_pkg\_create/process()}, and \texttt{coap\_response\_pkg\_create/process()} are provided by the \gls{coap} library. The \texttt{send()} and \texttt{receive()} functions are provided by the OS. For example Zephyr OS which is the OS we used for testing our libraries, provides standard BSD sockets \cite{bsdsoc}, which allow sending and receiving data over UDP or TCP.

The function \texttt{oscore\_init()} derives the \textit{Common IV}, \textit{Sender Key} and \textit{Recipient Key} from the \textit{Master Secret} using an HKDF function. We use a callback function to an implementation of the HKDF function that is provided by the user.

\begin{figure}[t] 
    \centering
    \includegraphics[width=1\linewidth]{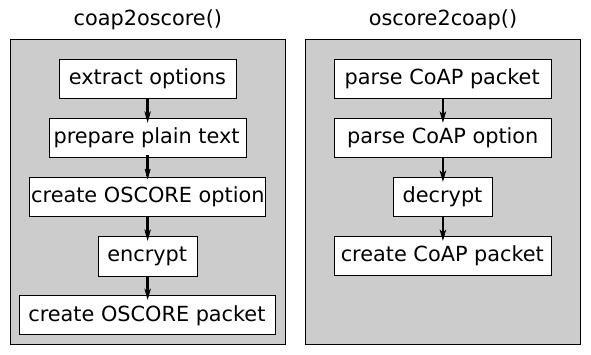}
    \caption{\texttt{coap2oscore()} and \texttt{oscore2coap()} conversion functions}
    \label{fig:oscoreinternals}
\end{figure}

The conversion functions \texttt{coap2oscore()} and \texttt{oscore2coap()} execute a series of operations as shown in \Cref{fig:oscoreinternals}.
In the \texttt{coap2oscore()} function, first, the options contained in the \gls{coap} packet are extracted. 
Then the options that need to be protected (E options), the payload and the code field form the plain text of the new \gls{oscore} packet, see also \Cref{fig:coap-to-oscore}. 
After that, the \gls{oscore} option is created. When \texttt{coap2oscore()} is called on the client side the \gls{oscore} option caries parameters used for the \gls{aead} nonce generation. Otherwise, the \gls{oscore} option is empty. The parameters carried in the \gls{oscore} option are saved locally in a state variable.
After that, a nonce is created and the plaintext is encrypted. For the encryption, we use a callback function to a user-provided implementation of the \gls{aead} algorithm. 
Then the \gls{oscore} packet is created.

In the \texttt{oscore2coap()} function, the \gls{oscore} packet is parsed. If the packet does not contain an \gls{oscore} option, the packet is a regular \gls{coap} packet. In this case, \texttt{oscore2coap()} returns with a status code indicating to the caller that the received packet is a \gls{coap} packet and can be processed as usual. If the packet contains an \gls{oscore} option, it is an \gls{oscore} packet. Then, the \gls{oscore} option is parsed in order to retrieve the parameters for the nonce generation. Note that these parameters are only carried in the \gls{oscore} option when \texttt{oscore2coap()} is called on the server side, otherwise they are retrieved from the local state variable. Afterwards, the payload is decrypted using a callback function to a user-provided implementation of the \gls{aead} algorithm. Next, a \gls{coap} packet is created.

\subsection{\uoscoretee{} Design}
Before discussing the design of \uoscoretee{} in detail, we first state which assets need to be protected and what are the consequences of potential attacks on them.

\subsubsection{Sensitive Assets}
By analyzing the \gls{oscore} specification we identified the \textit{Master Secret}, \textit{Recipient Key} and \textit{Sender Key} as sensitive key material. Leaking these keys will make Man-in-the-Middle attacks at the protocol level possible. 
Also, sensitive are the \gls{aead} and HKDF routines, which use these keys. If an attacker has access to these routines, he may manipulate them in a way that they leak their keys.

\subsubsection{Firmware and Key Isolation}
\uoscoretee{} reuses the main part of the \uoscore{} design. However, several important differences exist. 
First, the \gls{oscore} contexts are split between the \gls{tee} and the non-\gls{tee} domains. For each \gls{oscore} endpoint, the \gls{tee} domain maintains a data structure consisting of \textit{ID Context}, \textit{Master Secret}, \textit{Recipient Key} and \textit{Sender Key}. The non-\gls{tee} domain holds the contexts as discussed in \Cref{sec:oscorefunctionality} excluding the \textit{Master Secret}, \textit{Recipient Key} and \textit{Sender Key}.
Second, a narrow interface between the \gls{tee} domain and the non-\gls{tee} domain is provided by two functions \texttt{tee\_hkdf()} and \texttt{tee\_aead()}.
These functions implement the key derivation function and the \gls{aead} algorithm respectively.
Instead of any sensitive keys, they take \textit{ID Context} as an argument.
When they are called, \textit{ID Context} is passed to the \gls{tee} domain, where it is used to retrieve the cryptographic keys corresponding to the contexts in the non-\gls{tee} domain. When the keys are retrieved, the corresponding cryptographic algorithms are executed. 
In the case of \texttt{tee\_hkdf()}, the output of the function is the \textit{Recipient Key} or the \textit{Sender Key}, which is stored in the \gls{tee} domain. In the case of \texttt{tee\_aead()}, the output of the function is either a cipher text or a plain text, which is returned in the non-\gls{tee} domain.

%% file: edhoc.tex
\section{EDHOC}
In this section we give an overview of the \gls{edhoc} protocol functionality. Then we present the design of \uedhoc{} and \uedhoctee{}.

\subsection{\gls{edhoc} Functionality}
\begin{figure}[t] 
    \centering
    \includegraphics[width=.6\linewidth]{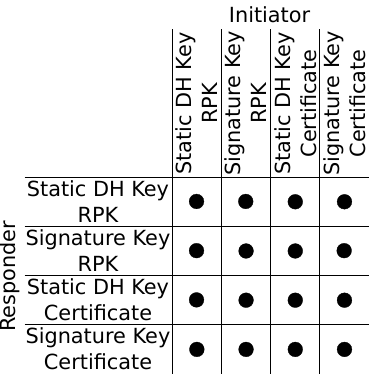}
    \caption{Possible combinations of initiator and responder authentication methods}
    \label{fig:authentication_method}
\end{figure}

\gls{edhoc} is a lightweight authenticated ephemeral \gls{dh} key exchange protocol \cite{edhoc} based on the SIGMA-I protocol \cite{Krawczyk2003}. 
\gls{edhoc} requires the exchange of three messages between an initiator and a responder endpoint. 
Additionally, each endpoint can indicate an error condition by sending an error message.
\gls{edhoc} can use static \gls{dh} keys or digital signatures for message authentication. Static \gls{dh} keys and digital signatures may be used with \glspl{rpk} or certificates. 
The authentication method used by the initiator may differ from the authentication method used by the responder, e.g., the initiator may use a static \gls{dh} key with \gls{rpk} and the responder may use asymmetric signature keys with certificate. In total, 16 different combinations are possible as shown in \Cref{fig:authentication_method}.
The main features of \gls{edhoc} are mutual authentication, \gls{pfs} and identity protection of the initiator and the responder against passive attackers. The identity of the initiator is also protected against active attackers.
On successful completion of the protocol, a pseudo random number $PRK$, depending on the \gls{dh} secret, and a transcript hash of the exchanged messages $TH$ are provided to the application, where a HKDF function referred in \cite{edhoc} as an \textit{exporter interface} is used for deriving application specific keys, e.g., an \gls{oscore} master secret.

\subsection{\uedhoc{} Design}
\uedhoc{} has two API functions, which the user needs to call on the initiator and responder side respectively. These functions are: \texttt{initiator\_run()} and \texttt{responder\_run()}. They successively execute the pre and post-processing of the \gls{edhoc} messages for all authentication modes. The differences in the protocol logic due to the different authentication modes are handled with different execution paths.
The \texttt{initiator\_run()} and \texttt{responder\_run()} functions take two sets of input parameters. The first set contains all endpoint-specific parameters, e.g., authentication keys, the endpoint's own certificate, etc. The second set of parameters consists of credentials belonging to potential communication peers, e.g., root public keys of \glspl{ca}.

When messages need to be sent or received inside \texttt{initiator\_run()} and \texttt{responder\_run()}, the special callback functions \texttt{tx()} and \texttt{rx()} are called.
The callback functions \texttt{tx()} and \texttt{rx()} must be implemented by the user. The implementation of those functions may use any protocol stack, thus our design is independent from it. 
Similarly, we use callback functions for all cryptographic operations which implementations must be provided by the user as well. The user may use a software crypto library or a hardware crypto accelerator, if one is available on the target microcontroller. In this way, we also achieve independence from the crypto implementation.

\uedhoc{} envisions a usage model, in which the user implements all required callback functions. The user then needs to initialize the radio peripherals to prepare them for sending and receiving, and afterwards call \texttt{initiator\_run()} on the initiator side and \texttt{responder\_run()} on the responder side. On successful protocol completion, \texttt{initiator\_run()} and \texttt{responder\_run()} return the required inputs for the exporter interface. 
If an error message is received during the execution of \texttt{initiator\_run()} and \texttt{responder\_run()} functions, they are aborted and the error message is returned to the caller.

\subsection{\uedhoctee{} Design}

Before we describe the design of \uedhoctee{} in detail, we provide an analysis of the sensitive assets that need to be protected, and describe the consequences of potential attacks on them.

\subsubsection{Sensitive Assets}
By analyzing the \gls{edhoc} specification, we identified the following key material as sensitive: 1) the long-term authentication keys, i.e., secret signature keys or secret static \gls{dh} keys, 2) the public authentication key of the other party when \gls{rpk} is used and the root public key of the \gls{ca} when certificates are used, 3) intermediate \gls{edhoc} keys and 4) the result of the \gls{edhoc} protocol. 
An attacker having access to one of those keys may: 1) impersonate the device by leaking its long-term authentication keys, 2) fool the device into believing that it is talking to a legitimate peer by changing the peer's public key or the root public key of the \gls{ca} and 3) compromise the security of \gls{edhoc} by leaking or manipulating its intermediate keys or the end result of the protocol. 

Also sensitive are all cryptographic routines using the keys. If these routines are manipulated, they may leak the keys or allow the \gls{edhoc} protocol to succeed although the communication peer is not authentic.

\subsubsection{Firmware and Key Isolation}
\uedhoctee{} reuses a main part of the \uedhoc{} design. In the following, we only discuss the relevant aspects of the key and crypto code isolation.

In the \gls{tee}, we use security contexts, which are data structures containing keys. Each device has at least two security context instances, one for its own keys, which we refer to as an \textit{own context}, and at least one for the credentials of the communication peers, which we refer to as \textit{peer context}. If certificates are used for the peer authentication, one peer context containing the root public key of the \gls{ca} is sufficient for authenticating many peers. However, when the peers authenticate with \gls{rpk}, or the peers use certificates issued by different \glspl{ca}, a separate context for each credential is used. 

Each security context is identified with a \textit{context ID}. Each key in a given context is identified with a \textit{key ID}. In our firmware design, crypto routines are called outside of the \gls{tee} by using the context and key IDs. Inside the \gls{tee}, the IDs are used to retrieve the keys. Then the crypto routine is executed with the retrieved key in the \gls{tee}. If the crypto routine produces new keys, e.g., if the crypto routine is a HKDF function, the new keys are stored in the own context of the device.

\uedhoctee{} uses the following interface functions between the \gls{tee} and the non-\gls{tee} domains: 
\texttt{aead()}, 
\texttt{asymm\_verify()}, 
\texttt{asymm\_sign()}, 
\texttt{hkdf\_extract()}, 
\texttt{hkdf\_expand()}, 
\texttt{dh\_secret\_derive()},  
\texttt{hash()} and \texttt{xor()}. 

Special attention is given to the functions \texttt{aead()} and \texttt{asymm\_verify()}, which verify the authenticity of a message. 
In the event that the authentication verification fails, these functions delete all intermediate keys, which prevents further protocol execution. 
If this is not done and the protocol logic outside the \gls{tee} is simply informed that the verification has failed, the protocol execution will not necessarily stop. A protocol logic outside the \gls{tee} controlled by an attacker will continue the protocol execution, which will lead to a situation where a key is exchanged with a non-authentic peer.

\section{Requirements and Design Goals Discussion}
Our designs and their implementations need to be suitable for constrained devices, see \textbf{R-I} in \Cref{sec:designgoals}. We address this by using the language C for our implementations and following the best practices for developing firmware for microcontrollers. In particular, we are not using dynamic memory allocation. All variables that we use are placed on the stack. The low memory requirements and low computational times of our implementation are demonstrated later in \Cref{sec:eval}.

Our designs need to ensure that the implementations can be transported to a variety of application environments, see \textbf{R-II} in \Cref{sec:designgoals}. 
Our usage models allow the independence of the embedded OS and the employed \gls{coap} library.
In order to be independent on the OS, our designs do not rely on any OS-specific functionalities.
In the case of \gls{oscore} in order to be independent of the \gls{coap} library, we integrated the minimal required subset of \gls{coap} message encoding and decoding functionalities (see \Cref{sec:oscore-coap conversion}) within our designs.
In the case of \gls{edhoc}, we achieve independence of the \gls{coap} library by using callback functions for to send and receive messages. We use the same strategy to achieve independence of the crypto implementation in all presented designs.

In order to minimize the chance of software attacks on the protocols (see \textbf{R-III} in \Cref{sec:designgoals}) we put all keys and the bare minimum code, consisting only of the crypto routines and minimal key handling, in a \gls{tee} domain. 
Additionally, \uoscoretee{} and \uedhoctee{} have a narrow interface between the \gls{tee} domain and the non-\gls{tee} domain. In the case of \uoscoretee{}, it only consists of two functions and in the case of \uedhoctee{}, eight. The parameters of all interface functions have primitive types. This allows us to avoid the need for complex parsers in the \gls{tee} domain, which are often prone to exploitable vulnerabilities.

%% file: evaluation.tex
\section{Evaluation}\label{sec:eval}
In this section we evaluate our implementations in terms of FLASH/RAM requirements, computational time and energy requirements.
For the evaluation of the protocols regarding their energy requirements we selected a communication stack using IPv6 over \gls{ble} as a representative low power IoT communication stack.

\subsection{Scope of the Evaluation}\label{sec:eval_scope}
In the fowling, we describe the scenarios in which we evaluate \gls{oscore} and \gls{edhoc}.
\subsubsection{\gls{oscore}}
Instead of trying to evaluate many scenarios with different payloads and combinations of protected and unprotected options, we decided to evaluate one representative scenario. 
This scenario comprises an \gls{oscore} server, which handles a complete message roundtrip -- 
the server receives an \gls{oscore} request, converts it to \gls{coap}, creates a new \gls{coap} response packet, converts it to an \gls{oscore} packet, and sends it. In our tests, we used small payloads typical for IoT applications. The \gls{oscore} request and the response messages are both 35\,byte. The corresponding \gls{coap} messages are 24\,byte.
\subsubsection{\gls{edhoc}}
\gls{edhoc} can use one of 16 combinations of initiator and responder authentication methods, see \Cref{fig:authentication_method}.
Our implementations are capable of handling all possible variants correctly. However, in this paper we present an evaluation of the modes in which both parties authenticate the same way: signatures with \gls{rpk}, signatures with certificate, static \gls{dh} keys with \gls{rpk} and static \gls{dh} keys with certificates, i.e., the modes on the diagonal in \Cref{fig:authentication_method}.
The RAM/FLASH, computational time and energy cost for the other modes can be estimated for the presented results. For example, if the initiator authenticates with a static \gls{dh} key with \gls{rpk} and the responder authenticates with a signature key with certificate the costs for this method will be higher than when both authenticate with \gls{dh} key with \gls{rpk} but lower than if both authenticate with signature key with certificate. 

In addition, in order to demonstrate the best possible \gls{edhoc} performance, we use native \gls{cbor} certificates as proposed in a current \gls{ietf} draft \cite{cborcert2020}. \gls{cbor} certificates are certificates in which the certificate information is \gls{cbor} encoded, thus these certificates can be parsed easily on constrained devices and are shorter in size. All \gls{cbor} certificates used in our evaluation are 135\,byte long. A summary of the \gls{edhoc} message sizes for the different authentication modes is given in \Cref{tab:edhoc_msg_size}.
\input{edhoc_message_sizes.tex}

\subsection{Evaluation Setups}
The RAM, FLASH and computational time evaluations of \uoscore{} and \uedhoc{} were conducted offline in a unit test environment where no messages ware sent. Those evaluations were performed on four widely used microcontroller architectures: Cortex M0, M4, M33 and Xtensa. This set of architectures covers a broad range of low-end CPU performance classes with CPU frequencies ranging from 16\,MHz to 160\,MHz.

To evaluate of the energy requirements of the protocols, we set up an IPv6 over \gls{ble} network consisting of an nRF52832 BLE SoC, a Raspberry Pi acting as a border router and a Linux workstation. We used BLE version 4.1 which allows 20\,byte of application payload data to be send in a single link layer packet. Data can be sent and received only during connection events in which it is possible to exchange several link layer packets per event. In our setting, the connection event interval was set to 50\,ms and the TX power was set to 0\,dBm.
In the case of \gls{oscore} on the nRF52832 SoC, we run an \gls{oscore} server, and on the Linux workstation we run a \gls{oscore} client. 
In the case of \gls{edhoc}, the energy requirements of the initiator and the responder were evaluated on the nRF52832 SoC in a series of experiments for the different authentication modes.

We evaluated \uoscoretee{} an \uedhoctee{} on the nRF9160 SoC from Nordic Semiconductors featuring a Cortex M33 CPU and a TrustZone-M extension. The nRF9160 SoC can be configured in two different ways: 1) the TrustZone is used, which means that all memories and peripherals are split in a secure and a non-secure world and 2) the TrustZone is not used, which means that the SoC behaves as regular microcontroller. We compare \uoscore{} and \uedhoc{} with \uoscoretee{} and \uedhoctee{} respectively by leveraging these two configuration modes. 

For our \gls{oscore} evaluation, we used the default algorithms AES-CCM-16-64-128 and HMAC-SHA256 as \gls{aead} and HKDF algorithm respectively. For our \gls{edhoc} evaluation, in addition to these algorithms we used c25519 and ed25519 as \gls{dh} and signature algorithms respectively. 

In order to present realistic results for the crypto operations, we used the crypto libraries tinycrypt \cite{tinycrypt} and c25519 \cite{c25519} which were developed especially for microcontrollers.
However, the performance and memory requirements of the crypto libraries (or hardware accelerator) have a high impact on the performance and memory requirements of the protocols.
In our evaluation we present the evaluation results related to the crypto library separately, thus practitioners interested in using some of the other popular crypto libraries for embedded devices may easily estimate the parameters of our protocols with a different crypto library. 
We refer users interested in dong so to the detailed evaluations of, e.g., wolfcrypt \cite{wolfcryptPerfomance}, mbedTLS \cite{mbedtlsPerfomance} and \textmu NaCl \cite{Dull2015}. 

For all experiments, we used a GCC compiler and set the optimization to -Os (optimized size). 

For the RAM evaluation, we measured the stack usage by filling the stack with a known pattern (0xaa) at boot time and checking how much of the pattern was overwritten just before and after the protocols ware executed. 

For all setups, we used the embedded OS Zephyr OS \cite{zephyr} on the microcontrollers with the Zephyr's \gls{coap} library \cite{zephyrCoAP}. On the Linux workstation, we used the \gls{coap} library cantcoap \cite{cantcoap}.

\subsection{\gls{oscore} Evaluation}
In the following, we present the \uoscore{} and \uoscoretee{} evaluation results.

\subsubsection{\uoscore{} FLASH, RAM, Computing Time and Energy Requirements}\label{sec:uoscore_eval}
\begin{figure}[h] 
    \centering
    \includegraphics[width=1\linewidth]{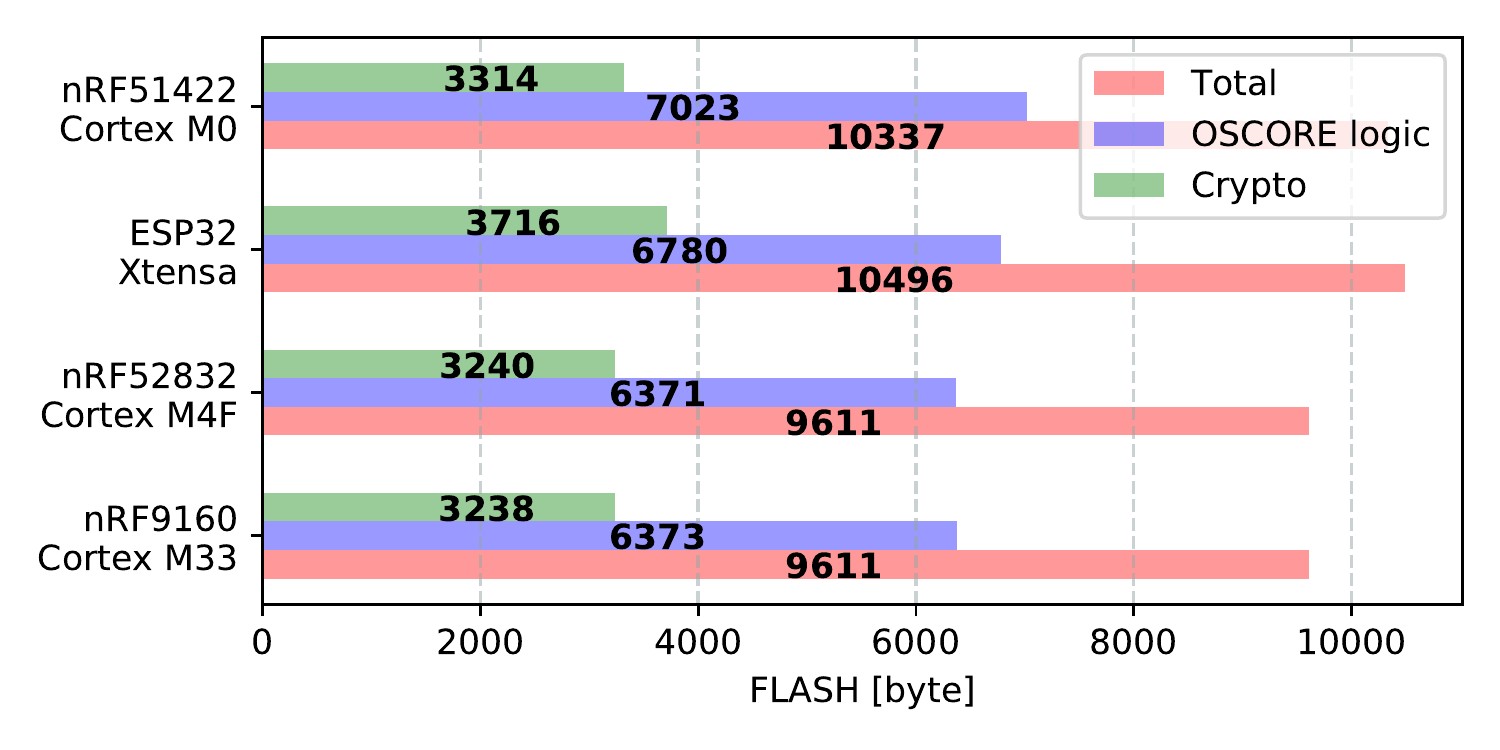}
    \caption{\uoscore{} FLASH requirements}
    \label{fig:oscore_flash}
\end{figure}

\begin{figure}[h] 
    \centering
    \includegraphics[width=0.7\linewidth]{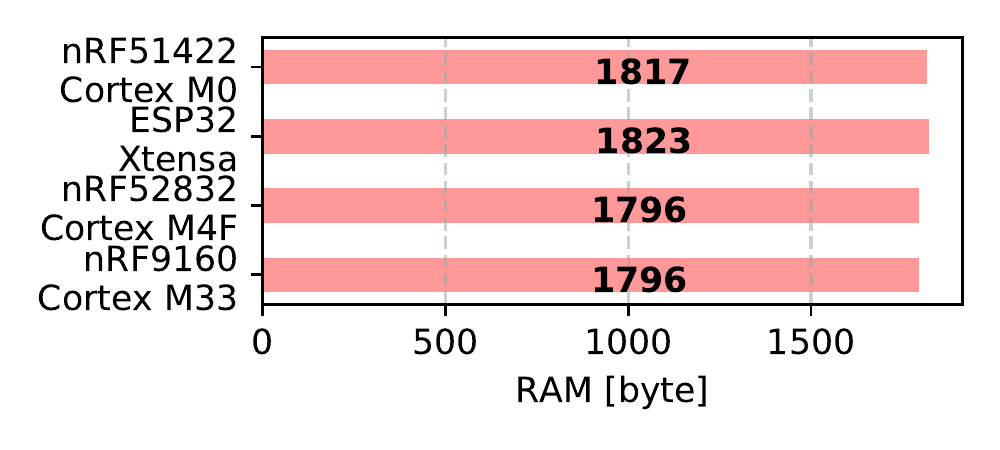}
    \caption{\uoscore{} RAM requirements}
    \label{fig:oscore_ram}
\end{figure}

The \uoscore{} FLASH and RAM requirements for the four different Cortex M0, M4, M33 and Xtensa architectures are shown in \Cref{fig:oscore_flash} and \Cref{fig:oscore_ram} respectively.
The FLASH requirements are split between requirements for the \gls{oscore} logic and requirements for the cryptographic operations. 
\uoscore{} requires in total $\approx$ 10\,KB FLASH and $\approx$ 1.8\,KB RAM. Approximately 6.3-7.0\,KB FLASH are required for the \gls{oscore} logic and 3.5\,KB for the crypto library. 
The FLASH requirements for the crypto library are very small since \gls{oscore} requires only AES-CCM and HMAC functions. The FLASH requirements for the crypto library can even be completely eliminated if a hardware crypto accelerator is used, thus the FLASH size can be reduced to $\approx$ 6.3-7.0\,KB.
Furthermore, the figures show that the differences due to the different CPU instruction sets are very small. 
In summary, the \uoscore{} FLASH and RAM requirements are very small, thus \uoscore{} is very well suited for constrained microcontrollers. 

\begin{figure}[h] 
    \centering
    \includegraphics[width=1\linewidth]{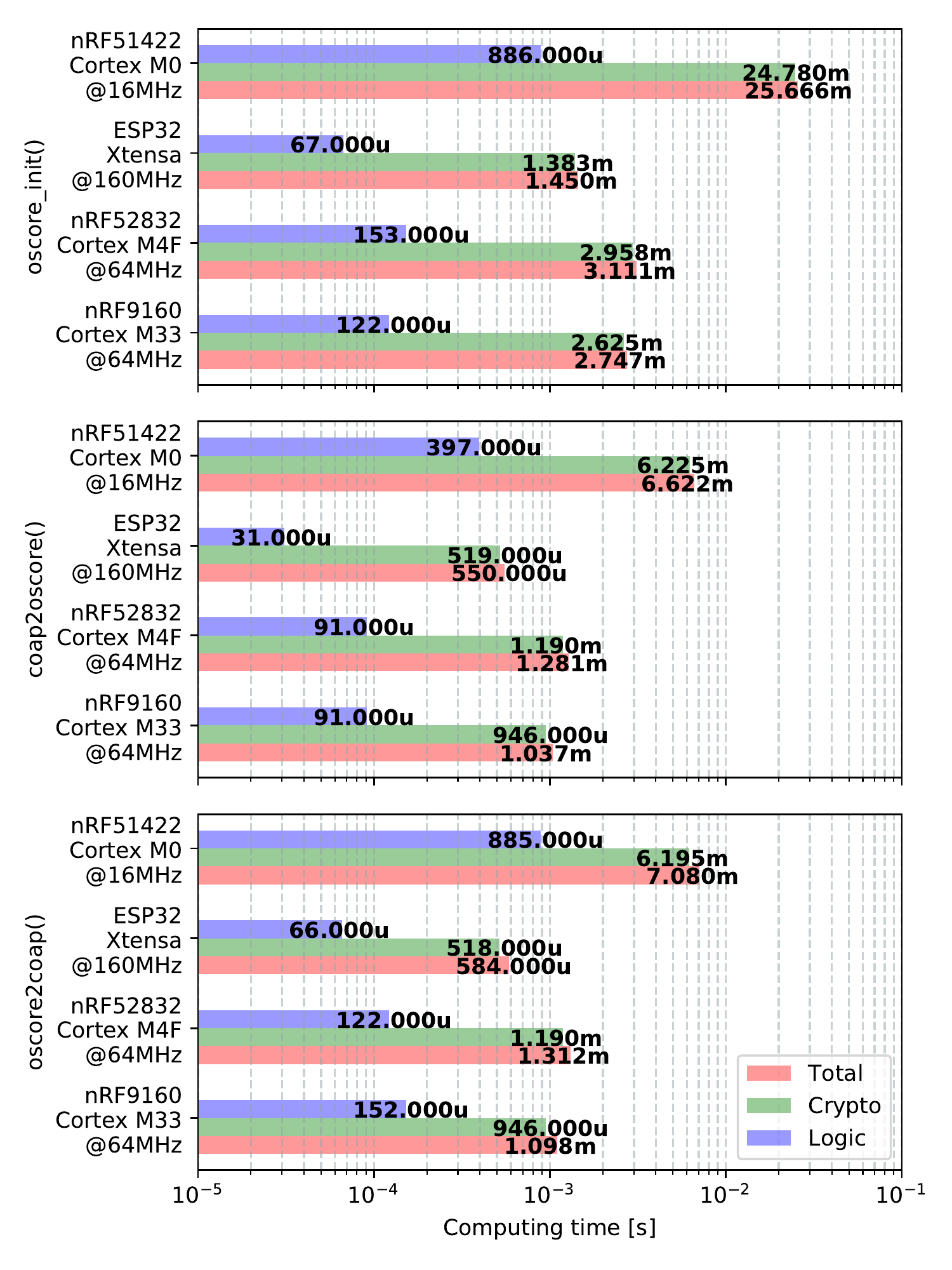}
    \caption{\uoscore{} computing time }
    \label{fig:oscore_comp_time}
\end{figure}

The \uoscore{} computing time measurements are given in \Cref{fig:oscore_comp_time}.
\Cref{fig:oscore_comp_time} shows that the most time-consuming function is the \texttt{oscore\_init()} function. However, this function needs to be executed only once in order to set up the sender and recipient contexts. During the regular \gls{oscore} operation, only the functions \texttt{coap2oscore()} and \texttt{oscore2coap()} are used. \Cref{fig:oscore_comp_time} shows also that, depending on the CPU architecture and clock frequency, the conversion functions convert a 35\,byte \gls{oscore} packet to \gls{coap} and vice versa in between 500\,\textmu s and few milliseconds, which shows that \uoscore{} is suitable for constrained microcontrollers also regarding computing time.

\begin{figure}[h] 
    \centering
    \includegraphics[width=1\linewidth]{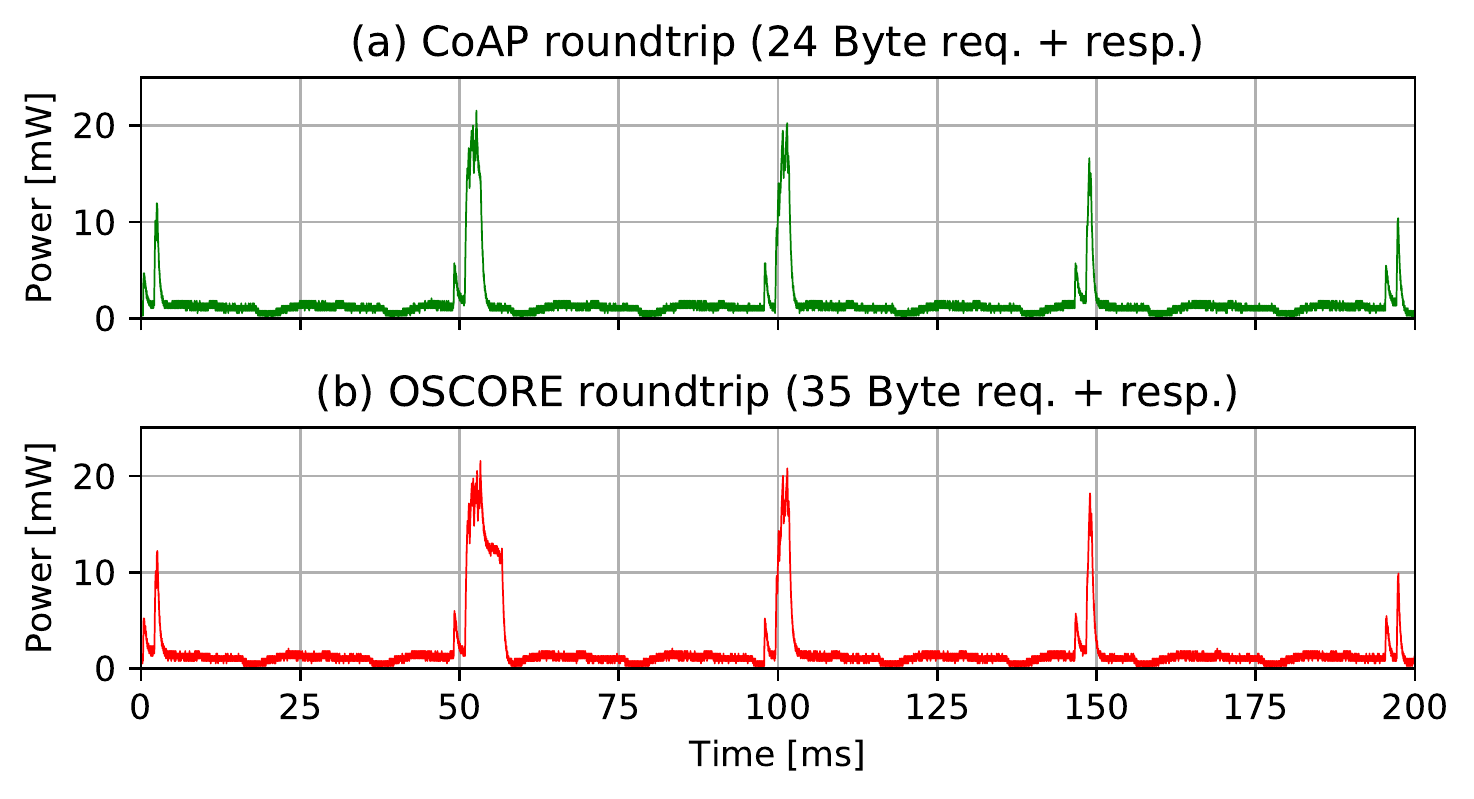}
    \caption{\gls{coap} and \gls{oscore} power consumption on a nRF52832 SoC in IPv6 over \gls{ble} network with connection event interval of 50\,ms}.
    \label{fig:oscore_power}
\end{figure}

We measured the power consumption of \uoscore{} running an nRF52832 \gls{ble} SoC in an IPv6 over \gls{ble} network while conducting two experiments. 
In the first experiment, we received and sent unprotected \gls{coap} packets 24\,byte in length. The power consumption of the device in this case is given in \Cref{fig:oscore_power}.(a).
In the second experiment, the nRF52832 SoC received a 35\,byte \gls{oscore} request then the request was converted to \gls{coap}. Afterwards, a response \gls{coap} packet was constructed and converted to \gls{oscore}, then sent back to the client, see \Cref{fig:oscore_power}.(b). 
In \Cref{fig:oscore_power} we see spikes in the power consumption of the SoC every 50\,ms. This is due to the \gls{ble} connection events during which the radio is active and sends and receives data. 
By integrating the measurements in \Cref{fig:oscore_power} over the time we calculated that the \gls{coap} roundtrip requires 352.97\,\textmu J and the equivalent \gls{oscore} protected roundtrip requires 366.63\,\textmu J. The \uoscore{} overhead is therefore only 3.87\%, 
which means that \uoscore{} is well suited for battery-powered devices. 
This means also that the \gls{oscore} energy costs are mainly (96.13\%) due to the radio transmission.
%

\subsubsection{\uoscoretee{} FLASH, RAM and Computing Time} 
\begin{figure}[h] 
    \centering
    \includegraphics[width=1\linewidth]{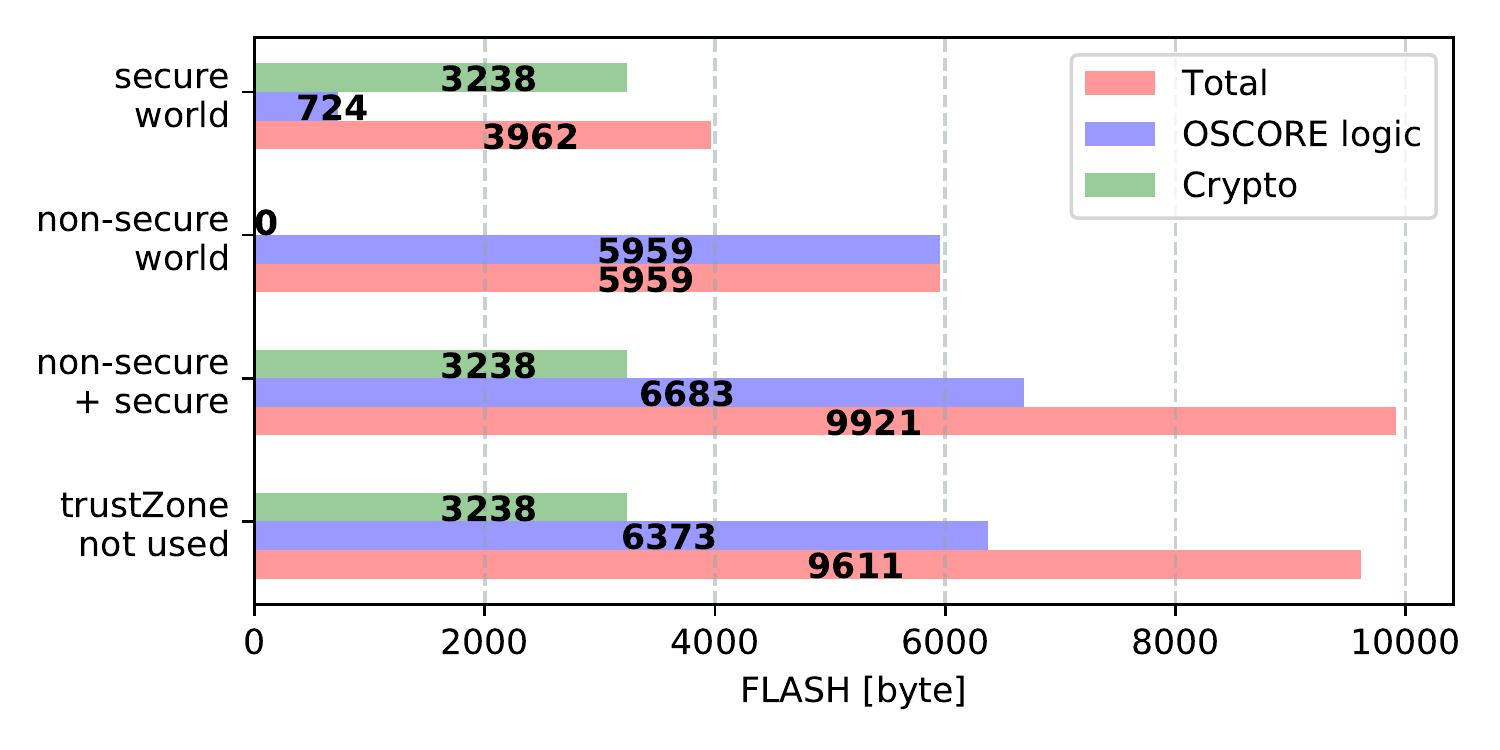}
    \caption{\gls{oscore} FLASH requirements when TrustZone is used vs. when TrustZone is not used}
    \label{fig:tz_oscore_flash}
\end{figure}

\Cref{fig:tz_oscore_flash} shows the FLASH sizes of the split between the secure and the non-secure world \uoscoretee{} implementation on an nrf9160 SoC. 
Additionally the FLASH size of the \uoscore{} when the TrustZone is not used is given for comparison.  
The code in the secure world is reduced to 3,962\,byte where the crypto library is 3,238\,byte and the \gls{oscore} logic is 724\,byte. On the other hand the non-secure world contains only \gls{oscore} logic code. In total \uoscoretee{} requires 9,921\,byte on a Cortex M33 CPU which is 3.13\% more than \uoscore{} which requires 9,611\,byte.

\begin{figure}[h] 
    \centering
    \includegraphics[width=0.7\linewidth]{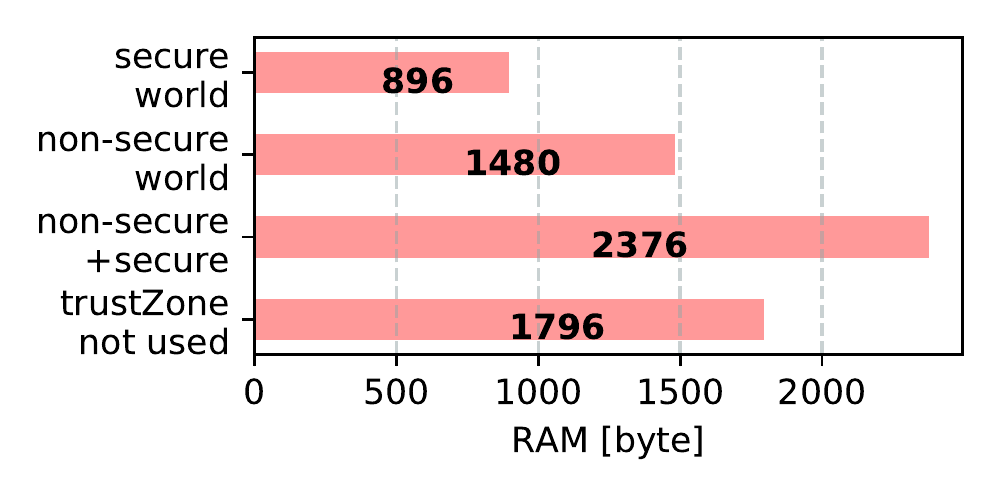}
    \caption{\gls{oscore} RAM requirements when TrustZone is used vs. when TrustZone is not used}
    \label{fig:tz_oscore_ram}
\end{figure}

\Cref{fig:tz_oscore_ram} shows the RAM requirements for the secure and non-secure world stacks, as well as the sum of both, and the stack requirements of \uoscore{}. 
In total, the secure and non-secure stack require 2,376\,byte which is 33.18\% more than \uoscore{}, which requires 1,784\,byte. This is because the secure world and the non-secure world have separated stacks, so memory cannot be reused as efficiently as when the complete implementation runs with a single stack. 
Although this increase appears extreme, the amount of RAM required by \uoscoretee{} is very low in the context of the total 256\,KB RAM available on the nRF9160 SoC.

\begin{figure}[h] 
    \centering
    \includegraphics[width=1\linewidth]{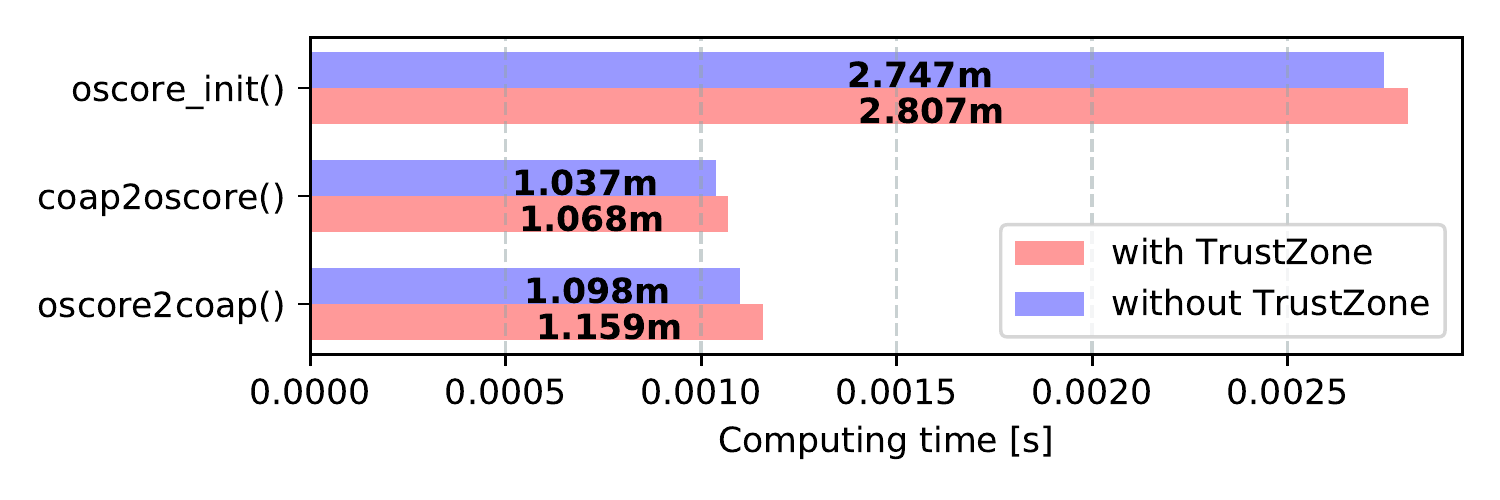}
    \caption{\uoscore{} and \uoscoretee{} computing time on nRF9160}
    \label{fig:tz_oscore_comp_time}
\end{figure}

\Cref{fig:tz_oscore_comp_time} shows the computing time of the three \gls{oscore} API functions for \uoscore{} and \uoscoretee{}.
The computing time overhead \uoscoretee{} is very low -- 3 to 5\,\% depending on the API function.
%

We provide no measurements of the energy consumption for \uoscoretee{} since we have shown that the \gls{oscore} energy costs are caused mainly by the radio transmissions (see energy evaluation for \uoscore{}), which depend only on the protocol stack and not on the \gls{oscore} library.

In summary, \uoscoretee{} is well suited for constrained devices as well, since it has very small overheads in terms of FLASH and computing time when compared to \uoscore{}. On the other hand, \uoscoretee{} has significantly higher RAM requirements, which, however, is still unproblematic for modern IoT SoCs such as nRF9160. 


\subsection{\gls{edhoc} Evaluation}
In the following, we present the \uedhoc{} and \uedhoctee{} evaluation results.

\subsubsection{\uedhoc{} FLASH, RAM, Computing Time and Energy Requirements}
\begin{figure}[h] 
    \centering
    \includegraphics[width=1\linewidth]{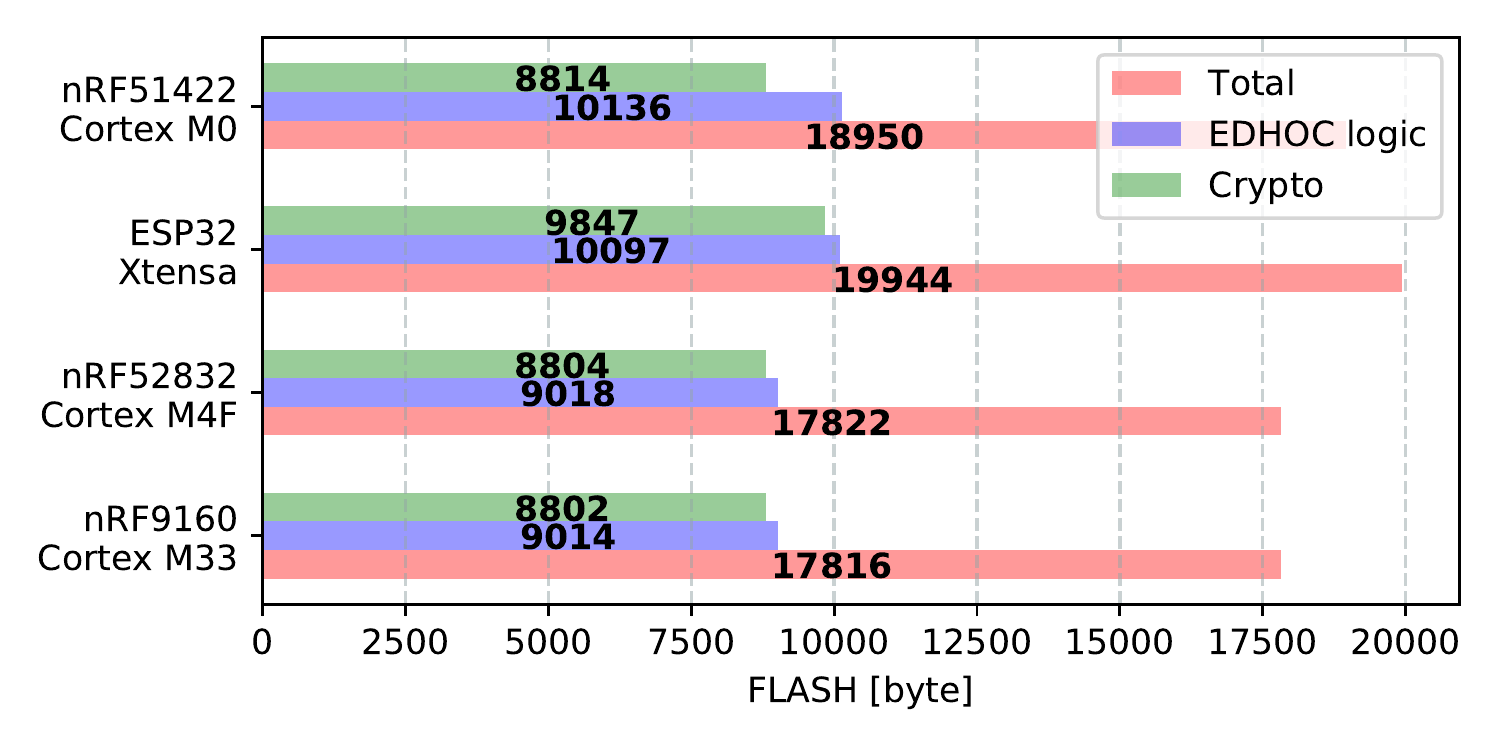}
    \caption{\uedhoc{} FLASH requirements}
    \label{fig:edhoc_flash}
\end{figure}

\begin{figure}[h] 
    \centering
    \includegraphics[width=1\linewidth]{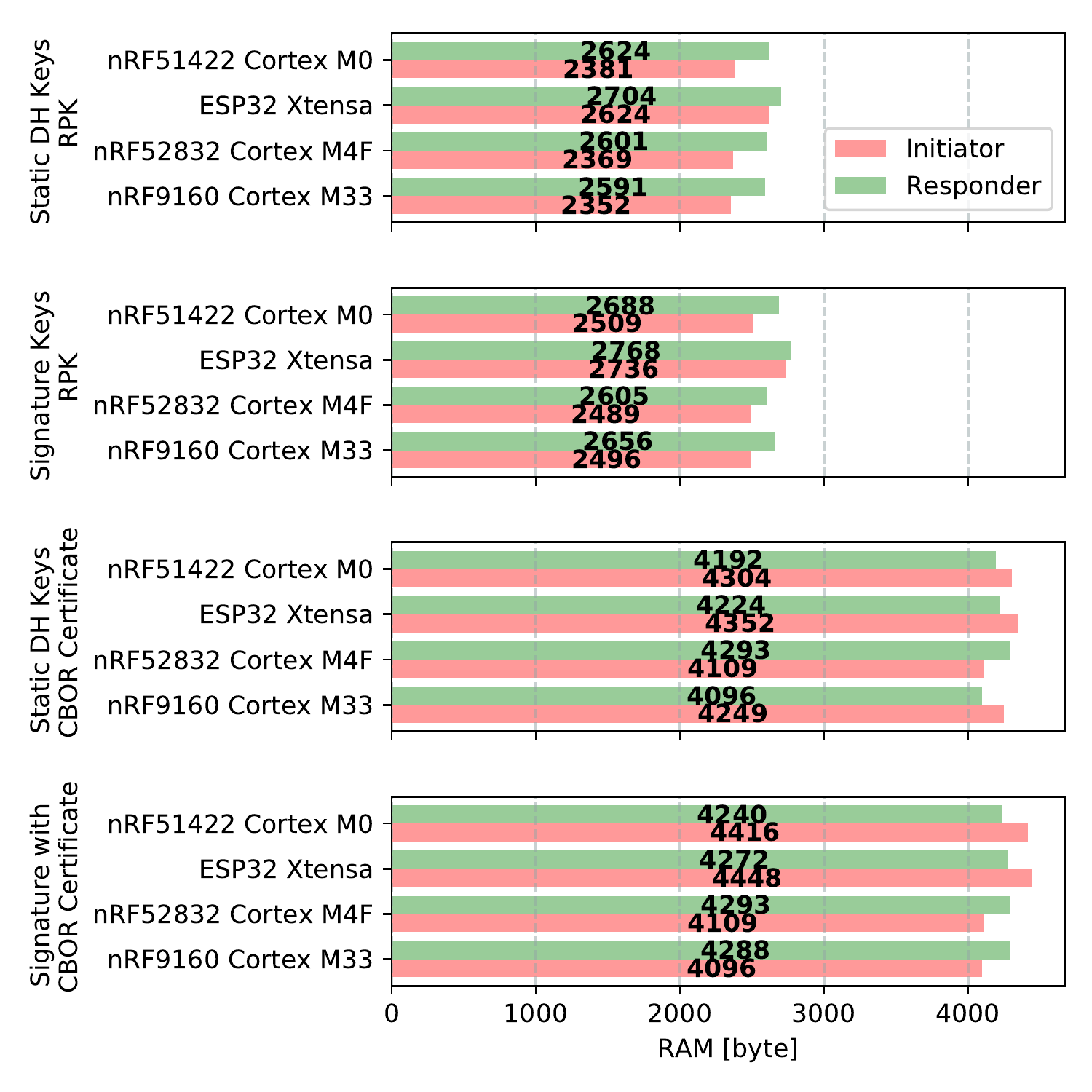}
    \caption{\uedhoc{} RAM requirements}
    \label{fig:edhoc_ram}
\end{figure}

The \uedhoc{} FLASH requirements are summarized in \Cref{fig:edhoc_flash}.
In total, \uedhoc{} requires between 17\,KB and 20\,KB depending on different CPU architectures, i.e., instruction sets. 
Approximately half of the FLASH is required for the protocol logic and the other half for the crypto library. 
The RAM requirements of \uedhoc{} are summarized in \Cref{fig:edhoc_ram}.
Depending on the authentication mode, \uedhoc{} requires between 2.4\,KB and 4.5\,KB RAM. The differences caused by the different CPU architectures are negligible. Initiator and Responder have also negligible differences. 
The differences in the different modes are due to the usage of memory buffers for the intermediary operations, which are bigger in the certificate modes and smaller in the \gls{rpk} modes.
In summary, the FLASH and RAM requirements of \uedhoc{} are reasonable for constrained microcontrollers. 


\begin{figure}[ht] 
    \centering
    \includegraphics[width=1\linewidth]{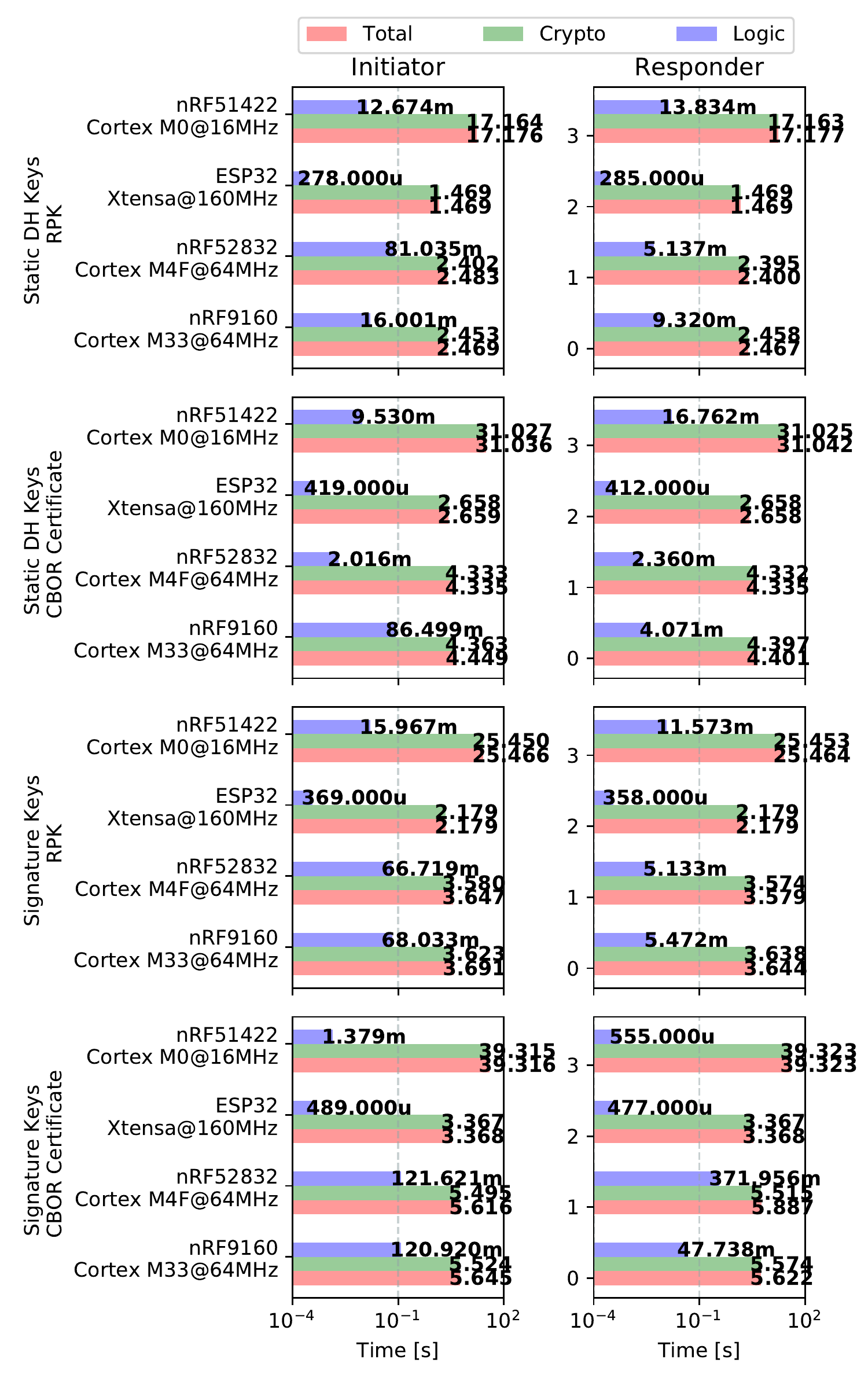}
    \caption{\uedhoc{} computing time}
    \label{fig:edhoc_copm_time}
\end{figure}

The \uedhoc{} computing time on the four different platforms is shown in \Cref{fig:edhoc_copm_time} (notice the logarithmic scale of the x axis).
The left part of the figure shows the computing times of the initiator and right part those of the responder.
In \Cref{fig:edhoc_copm_time}, we see that the computing times of initiator and responder are roughly equal. The total computing time is mainly defined by the crypto operations that require several thousand times more time then the protocol logic itself. 

Authentication with static \gls{dh} keys with \gls{rpk} is 45-50\% faster than authentication with asymmetric signatures with \gls{rpk}. 
Authentication with static \gls{dh} keys with certificates is 25-30\% faster than authentication with asymmetric signatures with certificates.
The usage of static \gls{dh} key authentication with certificates causes 75-80\% overhead compared to static \gls{dh} key authentication with \glspl{rpk}.
The usage of asymmetric key authentication with certificates causes 50-55\% overhead compared to asymmetric key authentication with \glspl{rpk}.
In \Cref{fig:edhoc_copm_time}, we can also see that on weaker devices such as nRF51422 (Cortex M0 running at 16\,MHz) the asymmetric authentication modes may be prohibitive for some applications since they require between $\approx$17 and $\approx$39 seconds. For devices that are more powerful, such as ESP32, nRF52832 and nRF9160, the computing times are lower and therefore acceptable for wider range of IoT applications.

\begin{figure}[h] 
    \centering
    \includegraphics[width=1\linewidth]{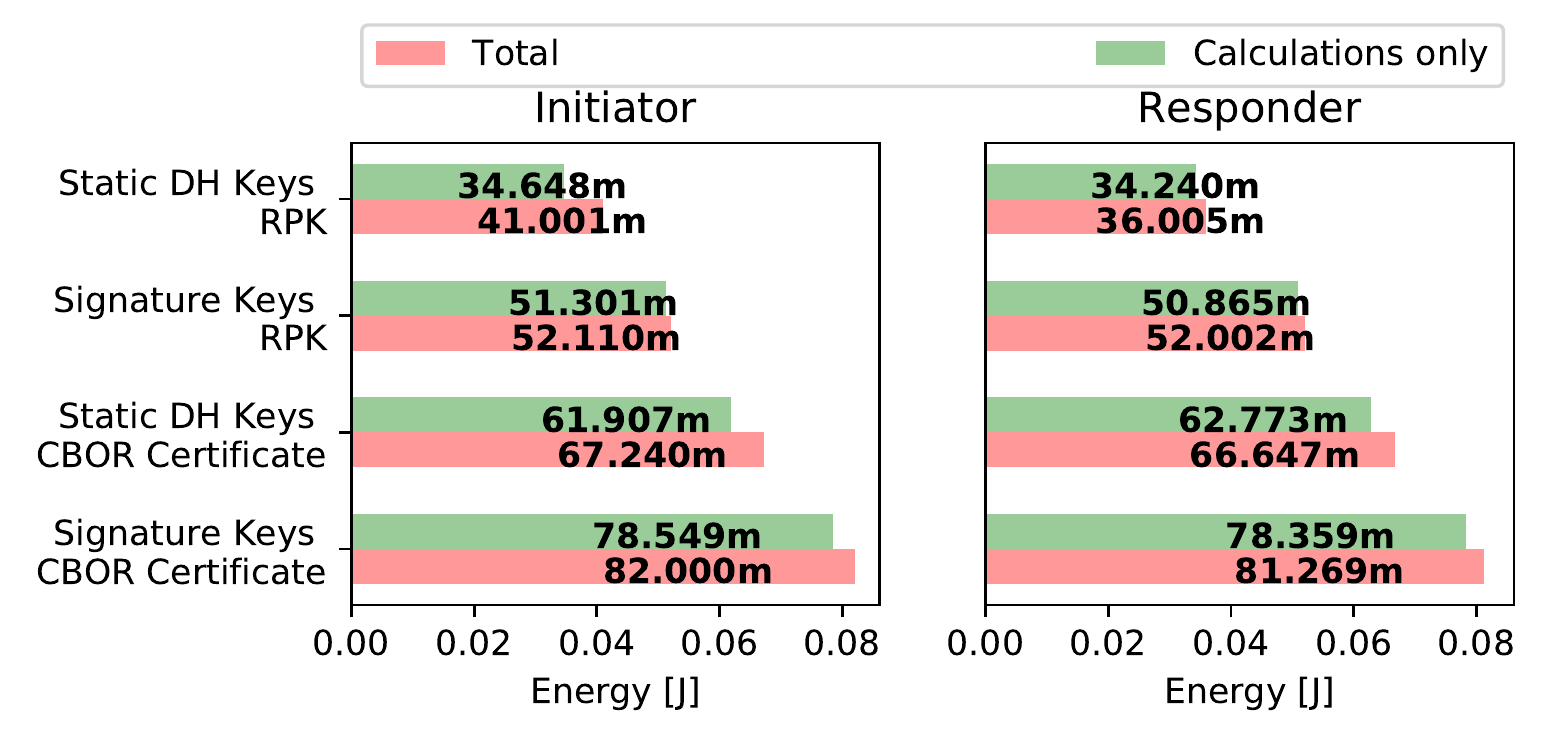}
    \caption{\uedhoc{} energy requirements on an nRF52832 SoC in IPv6 over \gls{ble} network}
    \label{fig:edhoc_power} 
\end{figure}

The \uedhoc{} energy costs for the initiator and responder in an IPv6 over \gls{ble} network are summarized in \Cref{fig:edhoc_power}. The energy costs are split into energy required for the calculation and total energy required for the complete protocol run, including the sending and receiving of messages.
In \Cref{fig:edhoc_power}, we see that the main part of the energy required is spent on the calculations. 

In order to estimate whether these numbers are prohibitive for some battery power application, we considered an example in which we assumed that an IoT device is powered by a typical CR-2032 coin cell battery with a capacity of 2,322\,J \cite{varta}. The energy capacity of 2,322\,J allows $\approx$28,000 to $\approx$67,000 protocol runs, depending on the authentication mode, therefore \uedhoc{} is well suited for the majority of battery-powered IoT devices.   

\subsection{\uedhoctee{} FLASH, RAM and Computing Time}
\begin{figure}[h] 
    \centering
    \includegraphics[width=1\linewidth]{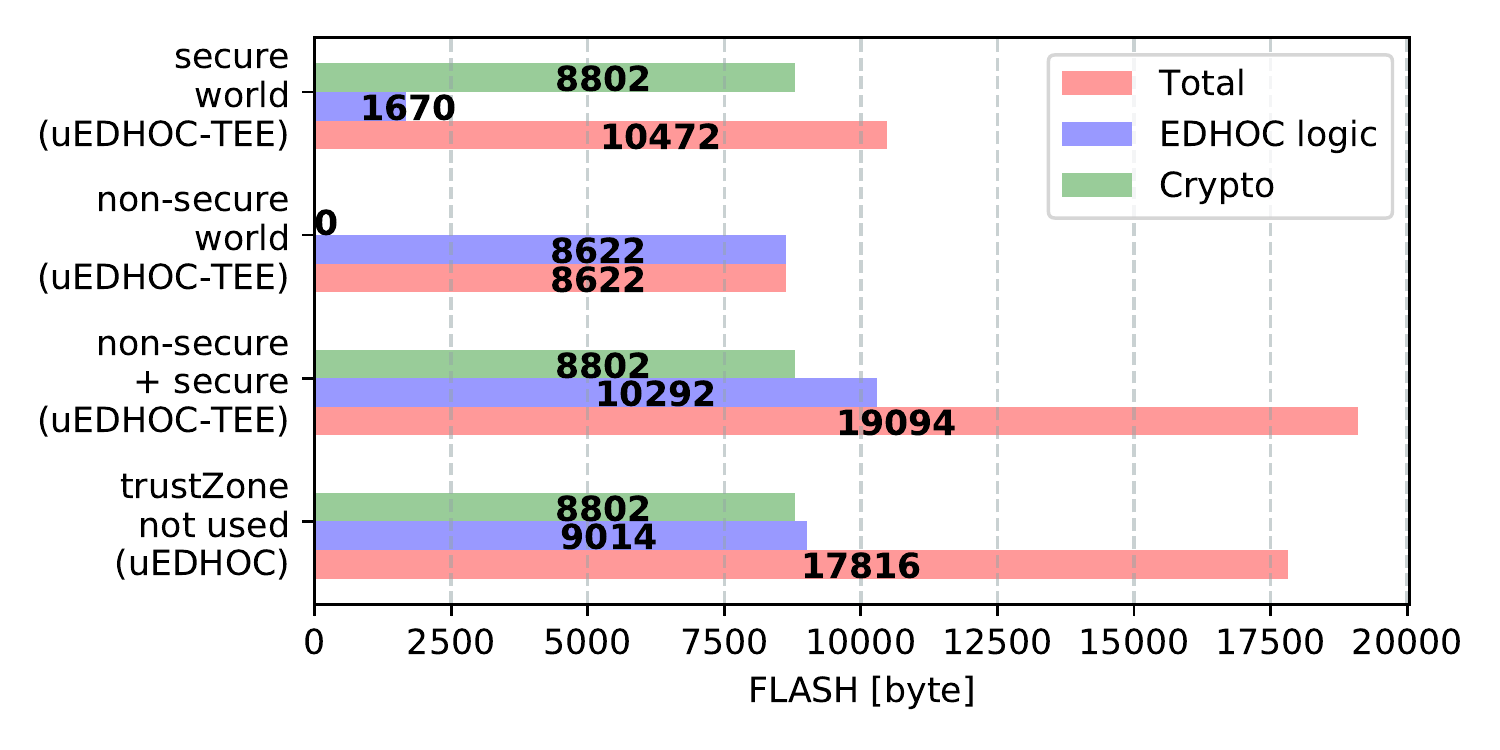}
    \caption{\gls{edhoc} FLASH requirements when TrustZone is used vs. TrustZone is not used}
    \label{fig:edhoc_tz_flash}
\end{figure}

\uedhoctee{} FLASH requirements for the secure and non-secure worlds are given in \Cref{fig:edhoc_tz_flash}. In \Cref{fig:edhoc_tz_flash}, we see that \uedhoctee{} requires in total 7.17\% more FLASH than \uedhoc{} . 
The percentual increase between the TrustZone and the non-TrustZone implementations of 7.17\% is higher as the equivalent percentual increase of 3.13\% for \gls{oscore}. This is because \uedhoctee{} contains additional code in the \gls{tee} domain for handling the more complex security contexts.

We measured the RAM and computational time requirements of \uedhoctee{} in the different authentication modes. The RAM overhead of \uedhoctee{}  is 30-40\,\% and the computational time overhead is 1-2\,\% when compared to \uedhoc{}.

We provide no detailed energy measurements of \uedhoctee{} because we have shown on the one hand side that the energy costs of \uedhoc{} are caused mainly by the calculations, see \Cref{fig:edhoc_power}. On the other hand side the computational overhead of \uedhoctee{} compared to \uedhoc is only 1-2\,\%. Therefore, the \uedhoctee{} energy costs can easily be estimated by adding 1-2\,\% to the results in \Cref{fig:edhoc_power}.

In summary,  \uedhoctee{} is well suitable for constrained devices as well, because of its low FLASH, RAM, computing time and energy requirements.

\subsection{Combined \gls{oscore} and \gls{edhoc} Usage}
In this section, we discuss the case when \gls{edhoc} and \gls{oscore} are used in succession.
All consideration in the following apply for both combinations \uoscore{}/\uedhoc{} and \uoscoretee{}/\uedhoctee{}. However, we provide details only for the combination \uoscore{}/\uedhoc{}  for the sake of clear presentation in this paper.

\begin{figure}[h] 
    \centering
    \includegraphics[width=1\linewidth]{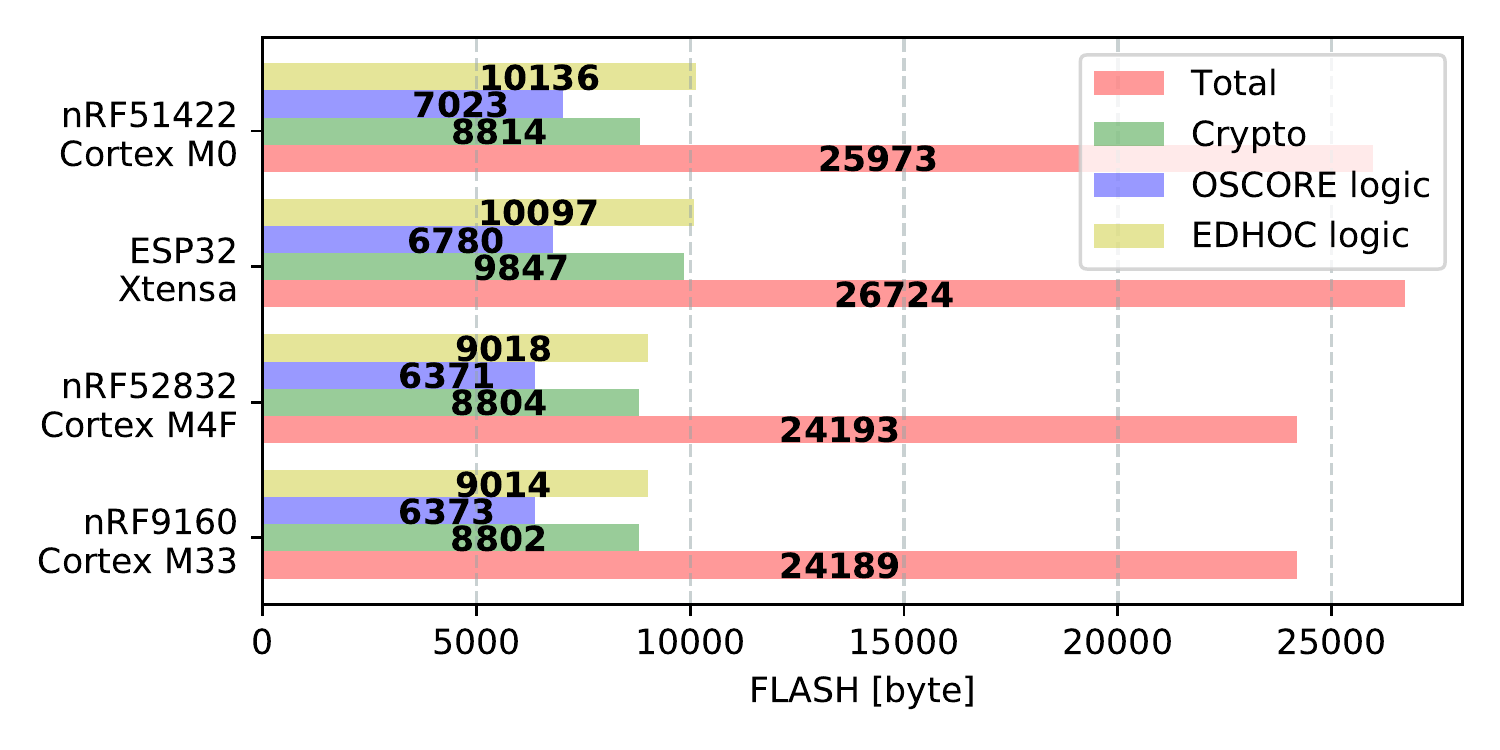}
    \caption{Combined \uoscore{} and \uedhoc{} FLASH requirements, when both are using the same \gls{aead} (CCM-16-64-128) and HKDF (HMAC-SHA256) algorithms.}
    \label{fig:combo_flash}
\end{figure}

The combined \uoscore{} and \uedhoc{} FLASH requirements are given in \Cref{fig:combo_flash}. The total FLASH requirements are the sum of the FLASH requirements for the protocols' logics and the crypto library. When \uoscore{} and \uedhoc{} are used with cipher suites relying on the same \gls{aead} and HKDF algorithms, e.g., both use AES-CCM-16-64-128 and HMAC-SHA256 the combined FLASH requirements for the crypto library are equal to the \uedhoc{} FLASH requirements for the crypto library.
\Cref{fig:combo_flash} shows that when \uoscore{} and \uedhoc{} are used together they require in total $\approx$25\,KB FLASH, which is reasonable for constrained microcontrollers.

When \uoscore{} and \uedhoc{} are used together, the total RAM requirements are determined by \uedhoc{}. This is due to the fact that we use only stack memory, which, once \uedhoc{} finishes its operations is reused by \uoscore{}. 

When the protocols are used in succession, the total computing time and energy requirements can be calculated as the sum of the individual protocols.

%% file: edhoc_message_sizes.tex

\begin{table}[!ht]
	\centering
	\begin{tabular}{cccc} 
		\toprule 
		 Authentication mode 							& Msg 1 & Msg 2 & Msg 3 \\
		\midrule 
		Static \gls{dh} keys / \gls{rpk} 	& 37 &  46 & 20\\
		Signature keys / \gls{rpk} 			& 37 & 117 & 91\\
		Static \gls{dh} keys / Certificate	& 37 & 186 & 160\\ 
		Signature keys / Certificate		& 37 & 243 & 217\\ 
		\bottomrule
	\end{tabular}
	\caption{\gls{edhoc} message sizes in bytes} 
	\label{tab:edhoc_msg_size}
\end{table} 

%% file: related_work.tex
\section{Related Work}\label{sec:related_work}
In this section, we give an overview of the related work in the area of \gls{oscore} and \gls{edhoc} evaluation and point out the differences to our paper.

\gls{oscore}, DTLS and NDN protocol evaluation in single and multi-hop scenarios are presented in \cite{Gndoan2020}. In this paper, the authors use the \gls{oscore} implementation libOSCORE \cite{liboscore}. The main difference between libOSCORE and \uoscore{} is that libOSCORE depends significantly on the \gls{coap} library \cite{liboscoreIntegration}. In contrast, our implementation contains the minimal required subset of \gls{coap} parsing/encoding functionalities. Additionally, it uses callbacks for the crypto operations. This makes our implementation completely independent of the \gls{coap} library, OS or crypto implementation (library or hardware accelerator). Moreover, \cite{Gndoan2020} concentrates mostly on the effects of packets lost in single and multi-hop scenarios, and does not present evaluation in terms of FLASH/RAM requirements, computing time and energy. Also \cite{Gndoan2020} does not consider the usage of a \gls{tee} for separating the security critical operations from the non-critical \gls{oscore} parsing and data encodings. 

In \cite{Perez2019}, an implementation and evaluation of \gls{edhoc} version 8 and an optimized variant of it relying on out-of-band parameter negotiation are presented. Both variants are evaluated only with \glspl{psk} and asymmetric signatures with \gls{rpk} authentication. Note that the \gls{edhoc} \gls{psk}  mode was removed from the later protocol drafts.
The runtime and success rates of both \gls{edhoc} variants are evaluated through simulation, considering packet losses in single and multi-hop environments. Also, the message size of both variants are compared with the \gls{dtls} 1.3 handshake.
In \cite{SanchezIborra2018}, the usage of \gls{edhoc} with \gls{psk} message authentication for updating session keys in LoRa networks is analyzed. The analysis concentrates on the message size overhead compared to \gls{dtls} and time-on-air. 
\cite{Perez2019a} presents an approach in which the \gls{edhoc}'s \glspl{psk} or \glspl{rpk} are derived by an LO-CoAP-EAP \cite{Carrillo2017} bootstrapping process. In addition, an \gls{edhoc} evaluation regarding message sizes and transmission times is presented. The paper considers only \gls{psk} authentication and asymmetric signature authentication with \glspl{rpk}. 
A detailed message size comparison between \gls{dtls} 1.2, \gls{dtls} 1.3, \gls{tls} 1.2, \gls{tls} 1.3, \gls{edhoc} and \gls{oscore} is provided by the \gls{ietf} in \cite{Mattsson2020}.
The previous work \cite{Perez2019, SanchezIborra2018, Perez2019a} in the area of \gls{edhoc} implementations lacks evaluation results of the more recently proposed authentication with static \gls{dh} keys. The usage of certificates is also not evaluated. Moreover, previous work lacks detailed evaluation regarding the FLASH/RAM requirements, computation times and energy consumption.
In contrast, we implement the latest version of \gls{edhoc}, taking into account all authentication modes.
For those, we provide a message size comparison and an extensive evaluation of the FLASH/RAM requirements, computation times and energy consumption on state-of-the-art Cortex M and Xtensa microcontrollers. In addition, we leverage the TrustZone \gls{tee} for executing all cryptographic operations and storing keys. We provide an evaluation of the overhead caused by this separation.

%% file: conclusion.tex
\section{Conclusion}
In this paper we presented the design of \uoscore{} and \uedhoc{} firmware libraries for constrained regular microcontrollers, which are based on the newest state of the \gls{oscore} and \gls{edhoc} specifications and consider all modes of operation. 
Additionally, we presented the design of \uoscoretee{} and \uedhoctee{} firmware libraries for microcontrollers featuring a \gls{tee}, which provide protection against attackers exploiting software vulnerabilities. This is achieved by separating the cryptographic keys and routines from the rest of the firmware, which may be vulnerable. 
We evaluated our libraries extensively on several broadly used microcontroller architectures. 
Our evaluation shows that when \uoscore{} and \uedhoc{} are used together they require a total of $\approx$25\,KB FLASH and between $\approx$1.8\,KB and  $\approx$4.2\,KB RAM depending on the \gls{edhoc} mode.
We also show that a typical \gls{coap} packet can be protected with \gls{oscore} within a few milliseconds. Our computing time evaluation of the \gls{edhoc} protocol shows that authentication with static \gls{dh} keys is 45-50\,\% faster than authentication with signatures, when \glspl{rpk} are used. When certificates are used, static \gls{dh} key authentication is 25-30\,\% faster than authentication with signatures. 
Our libraries for microcontrollers with a \gls{tee} show low overhead in terms of computing time and FLASH requirements. However, the RAM overhead is 30-40\,\%, which is still acceptable for the majority of IoT SoCs.


%% file: ms.bbl
\begin{thebibliography}{10}
\providecommand{\url}[1]{#1}
\csname url@samestyle\endcsname
\providecommand{\newblock}{\relax}
\providecommand{\bibinfo}[2]{#2}
\providecommand{\BIBentrySTDinterwordspacing}{\spaceskip=0pt\relax}
\providecommand{\BIBentryALTinterwordstretchfactor}{4}
\providecommand{\BIBentryALTinterwordspacing}{\spaceskip=\fontdimen2\font plus
\BIBentryALTinterwordstretchfactor\fontdimen3\font minus
  \fontdimen4\font\relax}
\providecommand{\BIBforeignlanguage}[2]{{%
\expandafter\ifx\csname l@#1\endcsname\relax
\typeout{** WARNING: IEEEtranS.bst: No hyphenation pattern has been}%
\typeout{** loaded for the language `#1'. Using the pattern for}%
\typeout{** the default language instead.}%
\else
\language=\csname l@#1\endcsname
\fi
#2}}
\providecommand{\BIBdecl}{\relax}
\BIBdecl

\bibitem{secureCode}
``{A few intricacies of writing Armv8-M Secure code},''
  \url{https://community.arm.com/developer/ip-products/processors/trustzone-for-armv8-m/b/blog/posts/a-few-intricacies-of-writing-armv8-m-secure-code},
  accessed: 2020-05-06.

\bibitem{wolfcryptPerfomance}
``{Benchmarking wolfSSL and wolfCRYPT},''
  \url{https://www.wolfssl.com/docs/benchmarks/}, accessed: 2020-09-22.

\bibitem{cantcoap}
``{cantcoap CoAP Library},'' \url{https://github.com/staropram/cantcoap},
  accessed: 2020-05-04.

\bibitem{varta}
``{CR 2032 Data Sheet},''
  \url{http://products.varta-microbattery.com/applications/MB_DATA/DOCUMENTS/DATA_SHEETS/DS6032.pdf},
  accessed: 2020-05-06.

\bibitem{liboscore}
``{libOSCORE: An OSCORE implementation (not only) for embedded systems},''
  \url{https://github.com/chrysn/liboscore}, accessed: 2020-04-23.

\bibitem{liboscoreIntegration}
``{libOSCORE levels of library integration},''
  \url{https://oscore.gitlab.io/liboscore/integration_levels.html}, accessed:
  2020-04-23.

\bibitem{LPC5500}
``{LPC5500 Series: World’s Arm® Cortex®-M33 based Microcontroller Series
  for Mass Market, Leveraging 40nm Embedded Flash Technology},''
  \url{https://www.nxp.com/products/processors-and-microcontrollers/arm-microcontrollers/general-purpose-mcus/lpc5500-cortex-m33:LPC5500_SERIES},
  accessed: 2020-05-06.

\bibitem{M2351}
``{M2351 Microcontroler from Nuvoton},'' \url{https://m2351.nuvoton.com/},
  accessed: 2020-05-02.

\bibitem{nrf91}
``{nRF9160 Low power SiP with integrated LTE-M/NB-IoT modem and GPS},''
  \url{https://www.nordicsemi.com/Products/Low-power-cellular-IoT/nRF9160},
  accessed: 2020-05-02.

\bibitem{STM32L55}
``{STM32L552RC Microcontroller},''
  \url{https://www.st.com/en/microcontrollers-microprocessors/stm32l552rc.html},
  accessed: 2020-05-06.

\bibitem{tinycrypt}
``{TinyCrypt Cryptographic Library},''
  \url{https://github.com/intel/tinycrypt/blob/master/documentation/tinycrypt.rst},
  accessed: 2020-05-06.

\bibitem{opensourcelink}
``{uOSCORE / uEDHOC},''
  \url{https://github.com/Fraunhofer-AISEC/uoscore-uedhoc}, accessed:
  2021-03-25.

\bibitem{heap}
``{What is Heap Fragmentation?}''
  \url{https://cpp4arduino.com/2018/11/06/what-is-heap-fragmentation.html},
  accessed: 2020-05-06.

\bibitem{zephyr}
``{Zephyr OS},'' \url{https://www.zephyrproject.org/}, accessed: 2020-05-04.

\bibitem{zephyrCoAP}
``{Zephyr OS CoAP Library},''
  \url{https://docs.zephyrproject.org/latest/reference/networking/coap.html?highlight=coap\#coap},
  accessed: 2020-05-04.

\bibitem{c25519}
D.~Beer, ``{Curve25519 and Ed25519 for low-memory systems},''
  \url{https://www.dlbeer.co.nz/oss/c25519.html}, accessed: 2020-05-06.

\bibitem{rfc7049}
\BIBentryALTinterwordspacing
C.~Bormann and P.~E. Hoffman, ``{Concise Binary Object Representation
  (CBOR)},'' RFC 7049, Oct. 2013. [Online]. Available:
  \url{https://rfc-editor.org/rfc/rfc7049.txt}
\BIBentrySTDinterwordspacing

\bibitem{rfc8323}
\BIBentryALTinterwordspacing
C.~Bormann, S.~Lemay, H.~Tschofenig, K.~Hartke, B.~Silverajan, and B.~Raymor,
  ``{CoAP (Constrained Application Protocol) over TCP, TLS, and WebSockets},''
  RFC 8323, Feb. 2018. [Online]. Available:
  \url{https://rfc-editor.org/rfc/rfc8323.txt}
\BIBentrySTDinterwordspacing

\bibitem{Castellani2012}
A.~P. {Castellani}, T.~{Fossati}, and S.~{Loreto}, ``{HTTP-CoAP Cross Protocol
  Proxy: an Implementation Viewpoint},'' in \emph{2012 IEEE 9th International
  Conference on Mobile Ad-Hoc and Sensor Systems (MASS 2012)}, vol. Supplement,
  Oct 2012, pp. 1--6.

\bibitem{rfc8075}
\BIBentryALTinterwordspacing
A.~P. Castellani, S.~Loreto, A.~Rahman, T.~Fossati, and E.~Dijk, ``{Guidelines
  for Mapping Implementations: HTTP to the Constrained Application Protocol
  (CoAP)},'' RFC 8075, Feb. 2017. [Online]. Available:
  \url{https://rfc-editor.org/rfc/rfc8075.txt}
\BIBentrySTDinterwordspacing

\bibitem{Dull2015}
M.~Düll, B.~Haase, G.~Hinterwälder, M.~Hutter, C.~Paar, A.~H. Sánchez, and
  P.~Schwabe, ``High-speed curve25519 on 8-bit, 16-bit, and 32-bit
  microcontrollers,'' Cryptology ePrint Archive, Report 2015/343, 2015,
  \url{https://eprint.iacr.org/2015/343}.

\bibitem{rfc7230}
\BIBentryALTinterwordspacing
R.~T. Fielding and J.~Reschke, ``{Hypertext Transfer Protocol (HTTP/1.1):
  Message Syntax and Routing},'' RFC 7230, Jun. 2014. [Online]. Available:
  \url{https://rfc-editor.org/rfc/rfc7230.txt}
\BIBentrySTDinterwordspacing

\bibitem{Fielding2000}
R.~T. Fielding, ``{Architectural Styles and the Design of Network-based
  Software Architectures},'' Ph.D. dissertation, 2000.

\bibitem{Carrillo2017}
D.~Garcia~Carrillo, R.~Marin-Lopez, A.~Kandasamy, and A.~Pelov, ``{A CoAP-based
  network access authentication service for low-power wide area networks:
  LO-CoAP-EAP},'' \emph{Sensors}, vol.~17, p. 2646, 11 2017.

\bibitem{Gndoan2020}
C.~{Gündoğan}, C.~{Amsüss}, T.~C. {Schmidt}, and M.~{Wählisch}, ``{IoT
  Content Object Security with OSCORE and NDN: A First Experimental
  Comparison},'' in \emph{2020 IFIP Networking Conference (Networking)}, 2020,
  pp. 19--27.

\bibitem{bsdsoc}
B.~Hall, ``{Beej’s Guide to Network Programming Using Internet Sockets},''
  \url{http://beej.us/guide/bgnet/pdf/bgnet_a4_c_1.pdf}, accessed: 2020-05-06.

\bibitem{Krawczyk2003}
H.~Krawczyk, ``{SIGMA: The 'SIGn-and-MAc'approach to authenticated
  Diffie-Hellman and its use in the IKE protocols},'' in \emph{Proceedings of
  Crypto'03}, vol. 2729, 08 2003, pp. 400--425.

\bibitem{rfc5869}
\BIBentryALTinterwordspacing
H.~Krawczyk and P.~Eronen, ``{HMAC-based Extract-and-Expand Key Derivation
  Function (HKDF)},'' RFC 5869, May 2010. [Online]. Available:
  \url{https://rfc-editor.org/rfc/rfc5869.txt}
\BIBentrySTDinterwordspacing

\bibitem{Lerche2012}
C.~{Lerche}, N.~{Laum}, F.~{Golatowski}, D.~{Timmermann}, and C.~{Niedermeier},
  ``{Connecting the Web with the Web of Things: Lessons Learned from
  Implementing a CoAP-HTTP Proxy},'' in \emph{2012 IEEE 9th International
  Conference on Mobile Ad-Hoc and Sensor Systems (MASS 2012)}, vol. Supplement,
  Oct 2012, pp. 1--8.

\bibitem{Lo2017}
\BIBentryALTinterwordspacing
O.~Lo, W.~J. Buchanan, and D.~Carson, ``Power analysis attacks on the aes-128
  s-box using differential power analysis (dpa) and correlation power analysis
  (cpa),'' \emph{Journal of Cyber Security Technology}, vol.~1, no.~2, pp.
  88--107, 2017. [Online]. Available:
  \url{https://doi.org/10.1080/23742917.2016.1231523}
\BIBentrySTDinterwordspacing

\bibitem{Ludovici2015}
A.~Ludovici and A.~Calveras, ``{A Proxy Design to Leverage the Interconnection
  of CoAP Wireless Sensor Networks with Web Applications},'' \emph{Sensors
  (Basel, Switzerland)}, vol.~15, pp. 1217--44, 01 2015.

\bibitem{Mattsson2020}
\BIBentryALTinterwordspacing
J.~Mattsson, F.~Palombini, and M.~Vučinić, ``{Comparison of CoAP Security
  Protocols},'' Internet Engineering Task Force, Internet-Draft
  draft-ietf-lwig-security-protocol-comparison-04, Mar. 2020, work in Progress.
  [Online]. Available:
  \url{https://datatracker.ietf.org/doc/html/draft-ietf-lwig-security-protocol-comparison-04}
\BIBentrySTDinterwordspacing

\bibitem{rfc7668}
\BIBentryALTinterwordspacing
J.~Nieminen, T.~Savolainen, M.~Isomaki, B.~Patil, Z.~Shelby, and C.~Gomez,
  ``{IPv6 over BLUETOOTH(R) Low Energy},'' RFC 7668, Oct. 2015. [Online].
  Available: \url{https://rfc-editor.org/rfc/rfc7668.txt}
\BIBentrySTDinterwordspacing

\bibitem{Obermaier2017}
J.~{Obermaier}, R.~{Specht}, and G.~{Sigl}, ``Fuzzy-glitch: A practical ring
  oscillator based clock glitch attack,'' in \emph{2017 International
  Conference on Applied Electronics (AE)}, 2017, pp. 1--6.

\bibitem{Obermaier2017a}
\BIBentryALTinterwordspacing
J.~Obermaier and S.~Tatschner, ``Shedding too much light on a
  microcontroller{\textquoteright}s firmware protection,'' in \emph{11th
  {USENIX} Workshop on Offensive Technologies ({WOOT} 17)}.\hskip 1em plus
  0.5em minus 0.4em\relax Vancouver, BC: {USENIX} Association, Aug. 2017.
  [Online]. Available:
  \url{https://www.usenix.org/conference/woot17/workshop-program/presentation/obermaier}
\BIBentrySTDinterwordspacing

\bibitem{Perez2019}
S.~{Pérez}, J.~L. {Hernández-Ramos}, S.~{Raza}, and A.~{Skarmeta},
  ``{Application Layer Key Establishment for End-to-End Security in IoT},''
  \emph{IEEE Internet of Things Journal}, pp. 1--1, 2019.

\bibitem{Perez2019a}
\BIBentryALTinterwordspacing
S.~Pérez, D.~Garcia-Carrillo, R.~Marín-López, J.~L. Hernández-Ramos,
  R.~Marín-Pérez, and A.~F. Skarmeta, ``{Architecture of Security Association
  Establishment Based on Bootstrapping Technologies for Enabling Secure IoT
  Infrastructures},'' \emph{Future Generation Computer Systems}, vol.~95, pp.
  570 -- 585, 2019. [Online]. Available:
  \url{http://www.sciencedirect.com/science/article/pii/S0167739X18325573}
\BIBentrySTDinterwordspacing

\bibitem{cborcert2020}
\BIBentryALTinterwordspacing
S.~Raza, J.~Höglund, G.~Selander, J.~Mattsson, and M.~Furuhed, ``{CBOR Profile
  of X.509 Certificates},'' Internet Engineering Task Force, Internet-Draft
  draft-raza-ace-cbor-certificates-04, Mar. 2020, work in Progress. [Online].
  Available:
  \url{https://datatracker.ietf.org/doc/html/draft-raza-ace-cbor-certificates-04}
\BIBentrySTDinterwordspacing

\bibitem{Rogaway2002}
\BIBentryALTinterwordspacing
P.~Rogaway, ``Authenticated-encryption with associated-data,'' in
  \emph{Proceedings of the 9th ACM Conference on Computer and Communications
  Security}, ser. CCS '02.\hskip 1em plus 0.5em minus 0.4em\relax New York, NY,
  USA: Association for Computing Machinery, 2002, p. 98–107. [Online].
  Available: \url{https://doi.org/10.1145/586110.586125}
\BIBentrySTDinterwordspacing

\bibitem{SanchezIborra2018}
R.~Sanchez-Iborra, J.~Sanchez-Gomez, S.~Pérez, P.~Fernandez, J.~Santa,
  J.~Hernández-Ramos, and A.~Skarmeta, ``{Enhancing LoRaWAN Security through a
  Lightweight and Authenticated Key Management Approach},'' \emph{Sensors},
  vol.~18, 06 2018.

\bibitem{rfc8152}
\BIBentryALTinterwordspacing
J.~Schaad, ``{CBOR Object Signing and Encryption (COSE)},'' RFC 8152, Jul.
  2017. [Online]. Available: \url{https://rfc-editor.org/rfc/rfc8152.txt}
\BIBentrySTDinterwordspacing

\bibitem{rfc8613}
\BIBentryALTinterwordspacing
G.~Selander, J.~Mattsson, F.~Palombini, and L.~Seitz, ``{Object Security for
  Constrained RESTful Environments (OSCORE)},'' RFC 8613, Jul. 2019. [Online].
  Available: \url{https://rfc-editor.org/rfc/rfc8613.txt}
\BIBentrySTDinterwordspacing

\bibitem{edhoc}
\BIBentryALTinterwordspacing
G.~Selander, J.~P. Mattsson, and F.~Palombini, ``{Ephemeral Diffie-Hellman Over
  COSE (EDHOC)},'' Internet Engineering Task Force, Internet-Draft
  draft-ietf-lake-edhoc-01, Aug. 2020, work in Progress. [Online]. Available:
  \url{https://datatracker.ietf.org/doc/html/draft-ietf-lake-edhoc-01}
\BIBentrySTDinterwordspacing

\bibitem{Selander2017}
\BIBentryALTinterwordspacing
G.~Selander, F.~Palombini, and K.~Hartke, ``{Requirements for CoAP End-To-End
  Security},'' Internet Engineering Task Force, Internet-Draft
  draft-hartke-core-e2e-security-reqs-03, Jul. 2017, work in Progress.
  [Online]. Available:
  \url{https://datatracker.ietf.org/doc/html/draft-hartke-core-e2e-security-reqs-03}
\BIBentrySTDinterwordspacing

\bibitem{rfc7252}
\BIBentryALTinterwordspacing
Z.~Shelby, K.~Hartke, and C.~Bormann, ``{The Constrained Application Protocol
  (CoAP)},'' RFC 7252, Jun. 2014. [Online]. Available:
  \url{https://rfc-editor.org/rfc/rfc7252.txt}
\BIBentrySTDinterwordspacing

\bibitem{Sulaeman2016}
A.~B. Sulaeman, F.~A. Ekadiyanto, and R.~F. Sari, ``{Performance Evaluation of
  HTTP-CoAP Proxy for Wireless Sensor and Actuator Networks},'' \emph{2016 IEEE
  Asia Pacific Conference on Wireless and Mobile (APWiMob)}, pp. 68--73, 2016.

\bibitem{mbedtlsPerfomance}
H.~{Tschofenig} and M.~{Pegourie-Gonnard}, ``{Performance of State-of-the-Art
  Cryptography on ARM-based Microprocessors},''
  \url{https://csrc.nist.gov/csrc/media/events/lightweight-cryptography-workshop-2015/documents/presentations/session7-vincent.pdf},
  accessed: 2020-09-22.

\bibitem{rfc7925}
\BIBentryALTinterwordspacing
H.~Tschofenig and T.~Fossati, ``{Transport Layer Security (TLS) / Datagram
  Transport Layer Security (DTLS) Profiles for the Internet of Things},'' RFC
  7925, Jul. 2016. [Online]. Available:
  \url{https://rfc-editor.org/rfc/rfc7925.txt}
\BIBentrySTDinterwordspacing

\bibitem{rfc3610}
\BIBentryALTinterwordspacing
D.~Whiting, R.~Housley, and N.~Ferguson, ``{Counter with CBC-MAC (CCM)},'' RFC
  3610, Sep. 2003. [Online]. Available:
  \url{https://rfc-editor.org/rfc/rfc3610.txt}
\BIBentrySTDinterwordspacing

\end{thebibliography}
